\documentclass[aps,prd,twocolumn,superscriptaddress,showpacs,amsmath,amssymb,
preprintnumbers,
nofootinbib,
]{revtex4-2}

\input{./definitions}

\begin{document}

\preprint{Belle~II Preprint 2024-006}
\preprint{KEK Preprint 2023-53}

\pacs{% WRITE THE PACS CODES IN THIS FILE.
}

\title{
New graph-neural-network flavor tagger for Belle~II and measurement of $\sin2\phi_1$ in  $B^0 \to J/\psi K^0_\text{S}$ decays
}
  \author{I.~Adachi\,\orcidlink{0000-0003-2287-0173}} % 2590
  \author{L.~Aggarwal\,\orcidlink{0000-0002-0909-7537}} % 10083
  \author{H.~Ahmed\,\orcidlink{0000-0003-3976-7498}} % 11323
  \author{H.~Aihara\,\orcidlink{0000-0002-1907-5964}} % 2223
  \author{N.~Akopov\,\orcidlink{0000-0002-4425-2096}} % 9443
  \author{A.~Aloisio\,\orcidlink{0000-0002-3883-6693}} % 2194
  \author{N.~Anh~Ky\,\orcidlink{0000-0003-0471-197X}} % 2218
  \author{D.~M.~Asner\,\orcidlink{0000-0002-1586-5790}} % 4684
  \author{H.~Atmacan\,\orcidlink{0000-0003-2435-501X}} % 2538
  \author{T.~Aushev\,\orcidlink{0000-0002-6347-7055}} % 3747
% ...
% 13403
\collaboration{The Belle II Collaboration}

\begin{abstract}
We present GFlaT, a new algorithm that uses a graph-neural-network to determine the flavor of neutral \PB mesons produced in \FourS decays. It improves previous algorithms by 
using the information from all charged final-state particles and the relations between them.
We evaluate its performance using \PB decays to flavor-specific hadronic final states reconstructed in a \SI{362}{fb^{-1}} sample of electron-positron collisions collected at the \FourS resonance with the Belle II detector at the SuperKEKB collider.
We achieve an effective tagging efficiency of \SI{37.40 \pm 0.43 \pm 0.36}{\percent},
where the first uncertainty is statistical and the second systematic, 
which is 18\% better than the previous Belle II algorithm.
Demonstrating the algorithm, we use $\PBzero\to\PJpsi\PKshortzero$ decays to measure the mixing-induced and direct \CP violation parameters, 
\SCP = \SI{0.724 \pm 0.035 \pm 0.009}{} and \CCP = \SI{-0.035 \pm 0.026 \pm 0.029}{}.

\end{abstract}

\maketitle

\newcommand{\prdcom}[1]{#1}

%%%%%%%%%%%%%%%%%%%%%%%%%%%%%%
\section{Introduction}

In the standard model, \CP violation arises from an irreducible complex phase in the Cabibbo-Kobayashi-Maskawa (CKM) matrix~\cite{Kobayashi:1973fv}.
Measurements of mixing-induced \CP violation in \PBzero meson decays constrain the values of the CKM-unitarity-triangle angles $\phi_1$ and $\phi_2$,\footnote{These angles are also known as $\beta$ and $\alpha$.}
helping us probe for sources of \CP violation beyond the standard model.
For example, we learn $\phi_1$ from $\PBzero\to\PJpsi\PKzero$~\cite{BaBar:2009byl,Belle:2012paq,LHCb:2023zcp} 
and $\phi_2$ from $\PBzero\to(\Ppi\Ppi)^0$~\cite{BaBar:2012fgk, Belle:2013epq, LHCb:2020byh}, $(\Prho\Prho)^0$~\cite{BaBar:2007cku, Belle:2015xfb, BaBar:2008xku}.
These measurements require knowledge of the neutral \PB meson flavor.
At \PB factory experiments, \PBzero and \APBzero mesons are produced in pairs from $\APelectron\Pelectron$ collisions at the \FourS resonance.
Since their states are entangled, tagging the flavor of one of the mesons, \PBtag, at the time of its decay determines the flavor of the other one, \PBsig, at the same time~\cite{Bigi:1981qs, Oddone:1987up}.

The \belletwo~\cite{Abe:2010gxa} experiment reported results using a flavor tagger~\cite{Belle-II:2021zvj,hadpaper,s2bpaper} based on algorithms developed by the \belle and \babar experiments~\cite{FTBelle, BaBar:2009byl}.
It uses the kinematic, topology, particle-identification, and charge information of charged final-state particles in the \PBtag decay
to infer if they originated from
categories of flavor-specific decays.
For instance, a charged particle is assigned as being a \APmuon in a $\PBzero\to D\APmuon\Pnum X$ decay or a \PKplus in the subsequent  $D\to\PKplus Y$ decay, the charge of which correlates to the \PBtag flavor.
This category-based flavor tagger selects the most probable assignment in each category, discards all other possibilities in that category, and then combines the probabilities of the selected assignments to predict the \PBtag flavor.

In this paper, we present a new algorithm, the graph-neural-network flavor tagger, \gflat, which uses a dynamic-graph-convolutional-neural-network~\cite{DGCNN} to combine the information from all charged final-state particles.
It improves flavor tagging by accounting for the discarded information in the category-based flavor tagger and correlations between information from final-state particles.

To demonstrate \gflat, we measure the \CP parameters of $\PBzero\to\PJpsi\PKshortzero$ from which we calculate $\phi_1$.
The probability density to observe \PBsig decay at a time $\Delta t$ from when \PBtag decays with flavor \qtag ($1$ for \PBzero, $-1$ for \APBzero) is
\begin{align}
 P(\Delta t, \qtag) = {} &\frac{e^{-\abs*{\Delta t}/\tau}}{4\tau} \big\{1 + \qtag (1 - 2w)[\SCP \sine(\Delta m_{\Pdown} \Delta t) \nonumber\\
 &~ - \CCP \cosine(\Delta m_{\Pdown} \Delta t)]\big\},
    \label{eq:dt_cp_perf}
\end{align}
where 
\qtag is determined by the flavor tagger, $w$ is the probability to wrongly determine it, 
$\tau$ is the \PBzero lifetime, 
and $\Delta m_{\Pdown}$ is the difference of masses of the \PBzero mass eigenstates.\footnote{We use a system of units in which $\hbar=c=1$ and mass and frequency have the same dimension.}
Here $S$ and $C$, the parameters of interest, quantify mixing-induced and direct \CP violation, respectively. In the standard model, $\SCP = \sin2\phi_1$ and $\CCP = 0$ to good precision~\cite{DeBruyn:2014oga, Barel:2020jvf, Barel:2022wfr}.
At \PB factories, the \PB mesons are boosted and have significant momentum in the lab frame, so $\Delta t$ is  determined from the relative displacement of their decay vertices.

To measure \CP parameters in tagged \PBzero decays, we must know $w$.
We determine it from events with the flavor-specific \PBsig decaying as $\PBzero\to\PDoptstarminus\Ppiplus$, for which
\begin{align}
    &P(\Delta t, \qsig, \qtag) = \nonumber\\
    &~ \frac{e^{-\abs*{\Delta t} / \tau}}{4\tau} \qty{1 - \qsig\qtag (1 - 2w) \cosine(\Delta m_{\Pdown} \Delta t)},
    \label{eq:dt_flav_perf}
\end{align}
where \qsig equals the charge of the pion from the \PBsig decay,
neglecting the $\mathcal{O}(\num{e-4})$ wrong-sign 
contribution from $\PBzero\to\PDoptstarplus\Ppiminus$~\cite{Dunietz:1997in, Belle:2010igk, BaBar:2008xvo};
we implicitly include charge conjugated decays here and throughout.
Here we assume $w$ is independent of the \PBsig decay mode.
Flavor taggers also determine the quality of their flavor assignments 
by the dilution factor, $r \in [0,1]$
which approximates $1 - 2w$.
We determine $w$ in seven contiguous disjoint intervals ($r$ bins) defined by the edges $[0.0, 0.1, 0.25, 0.45, 0.6, 0.725, 0.875, 1.0]$, as in Ref.~\cite{hadpaper}, and calculate the effective tagging efficiency,
\begin{equation}
    \efftag = \sum_i \effi (1 - 2 w_i)^2,
\end{equation}
\prdcom{where \effi is the efficiency for a \PB to be reconstructed in bin $i$}.
An increase in \efftag improves statistical precision for parameters measured in tagged \PBzero decays,
for example, the statistical uncertainties on \SCP and \CCP are proportional to $1/\sqrt{\efftag}$.
The effective tagging efficiency is thus a convenient metric for evaluating tagger performance.

We reconstruct the flavor-specific $\PBzero\to\PDoptstarminus\Ppiplus$ decays from $\PDminus\to\PKplus\Ppiminus\Ppiminus$ and  $\PDstarminus\to\APDzero\Ppiminus$ with $\APDzero\to\PKplus\Ppiminus$, $\PKplus\Ppiminus\Ppizero$, or $\PKplus\Ppiminus\Ppiplus\Ppiminus$.
We fit the background-subtracted $\Delta t$ distributions~\prdcom{\cite{Pivk:2004ty, Dembinski:2021kim}} to extract flavor tagger parameters, including $w$, and determine the $\Delta t$ resolution model.

For the measurements of \SCP and \CCP,
we reconstruct the \PBsig candidates by combining 
$\PKshortzero\to\Ppiplus\Ppiminus$ with $\PJpsi\to\APelectron\Pelectron$ or $\APmuon\Pmuon$.
The values of \SCP and \CCP are extracted via a fit to the background-subtracted $\Delta t$ distribution using the flavor tagger parameters and $\Delta t$ resolution model determined from the study of $\PBzero\to\PDoptstarminus\Ppiplus$.

This paper is organized as follows. We first discuss the Belle II detector and the simulation software used in the study in Sec.~\ref{sec:DetectorAndSimulation}. Section~\ref{sec:gflat} describes the \gflat algorithm, including input variables, training procedure, and a discussion on the improvement from the category-based flavor tagger. Section~\ref{sec:CalibrationAndPerformance} presents the evaluation of \gflat's performance using the flavor-specific process, $\PBzero\to\PDoptstarminus\Ppiplus$. We describe the measurement of \SCP and \CCP for $\PBzero\to\PJpsi\PKshortzero$ to demonstrate \gflat's effectiveness in Sec.~\ref{sec:MeasurementOfSin2phi1} and conclude in Sec.~\ref{sec:summary}.

%%%%%%%%%%%%%%%%%%%%%%%%%%%%%%%%%
\section{Detector and simulation}
\label{sec:DetectorAndSimulation}
We evaluate \gflat's performance using a \SI{362 \pm 2}{fb^{-1}} data set collected with the \belletwo detector in 2019--2022.
The \belletwo detector is located at SuperKEKB, which collides electrons and positrons at and near the \FourS resonance~\cite{Akai:2018mbz}.
It is cylindrical and includes a two-layer silicon-pixel detector~(PXD) surrounded by a four-layer double-sided silicon-strip detector~\cite{Belle-IISVD:2022upf} and a 56-layer central drift chamber~(CDC).
These detectors reconstruct 
trajectories of charged particles (tracks).
Only one sixth of the second layer of the PXD was installed for 
the data analyzed here.
The symmetry axis of these detectors, $z$, is nearly coincident with the direction of the electron beam.
Surrounding the CDC, which also measures \dedx ionization energy-loss, is a time-of-propagation 
detector~\cite{Kotchetkov:2018qzw} in the barrel and an aerogel-based ring-imaging Cherenkov detector
in the forward ($+z$) endcap region.
These detectors provide information for charged-particle identification.
Surrounding them is an electromagnetic calorimeter (ECL) based on CsI(Tl) crystals that primarily measures the energies and times of detection of photons and electrons.
Outside it is a superconducting solenoid magnet that provides a \SI{1.5}{T} field in the $z$ 
direction. Its flux return is instrumented with resistive-plate chambers and plastic scintillator modules to detect muons, \PKlongzero, and neutrons.

We use simulated data to train \gflat, estimate reconstruction efficiencies and background contributions, and construct fit models.
We generate $\APelectron\Pelectron\to\FourS\to\PB\APB$ using \EvtGen~\cite{Lange:2001uf} and \Pythia~\cite{Sjostrand:2014zea} and 
$\APelectron\Pelectron\to\Pquark\APquark$ with $\Pquark$ indicating a $u,d,c,$ or $s$ quark using \KKMC~\cite{Jadach:1999vf} and \Pythia.
We simulate particle decays using \EvtGen interfaced with \Pythia, and the interaction of particles with the detector using \Geant~\cite{Agostinelli:2002hh}.
Our simulation includes effects of beam-induced backgrounds~\cite{Lewis:2018ayu}.
Events in both simulation and data are reconstructed using the \belletwo analysis software framework~\cite{Kuhr:2018lps,basf2-zenodo}.

%%%%%%%%%%%%%%%%%%%%%%%%%%%%%%%
\section{\gflat}
\label{sec:gflat}

\gflat is designed to run after \PBsig is reconstructed and uses information from the tracks and energy deposits in the ECL (clusters) not associated with \PBsig, in the same manner as the category-based flavor tagger~\cite{Belle-II:2021zvj}.
We refer to these tracks and clusters as the rest of the event~(ROE), which mostly originates from \PBtag.
Tracks from the ROE must be within the CDC and have points of closest approach (POCAs) to the 
$\APelectron\Pelectron$ interaction region (IR) that are less than \SI{3}{cm} from the IR in the $z$ direction and less than \SI{1}{cm} from it in the transverse plane.
The shape and location of the IR are determined from $\APelectron\Pelectron\to\APmuon\Pmuon$ events in 30-minute intervals.
We retain only the first 16 charged particles in the ROE, ordered by decreasing momentum in the lab frame.
According to simulation, 
the average number of charged particles in the ROE is 4.8, and less than 0.001\% of events have more than 16 charged particles.

\gflat uses 25 input variables for each ROE charged particle:
the lab-frame Cartesian components of its momentum and the displacement of its POCA from the IR;
particle-identification likelihoods for each of the six possible charged final-state particles, \Pe, \Pmu, \Ppi, \PK, 
proton, and deuteron;
and the products of the charge of the particle and the output of the category-based flavor tagger for each of its 13 categories.\footnote{corresponding to $q_{\rm cand}y_{\rm cat}$ defined in Ref.~\cite{Belle-II:2021zvj}.}
The input variables have the same distributions for \PBzero and \APBzero except for 
differences in the detection and reconstruction efficiency for negative and positive charged particles.

\gflat uses a dynamic-graph-convolutional-neural-network that has been used for jet tagging at LHC experiments~\cite{Qu:2019gqs}.
\gflat first processes the input variables using the EdgeConv algorithm~\cite{DGCNN}, which consists of three neural networks: 
edge and node networks run in parallel, and a weight network runs on their output.
In the context of graph-neural-networks, the set of ROE charged particles is a graph with each particle a node and each pair an edge.
The node network processes the variables of each particle to update them. 
The edge network processes the variables of each pair of particles to update the variables of each particle.
To reduce computational resources, with no impact on performance, the edge network processes information from pairs formed from only the five nearest neighbors to each particle.
The weight network processes the outputs of the edge and node networks with a squeeze-and-excitation algorithm that calculates weights based on variable importance~\cite{SEblock}.
The output of the EdgeConv consists of the updated variables for each particle that are improved to more accurately reflect the characteristics of each particle.

\gflat runs EdgeConv twice.
The first run processes the measured particle variables, with its edge network finding nearest neighbors based on POCAs.
The second run processes the output of the first run, with its edge network finding nearest neighbors based on particle similarity using the updated particle variables.
To keep output reasonably symmetric between \PBzero and \APBzero, 
the output variables of each particle from the second EdgeConv are multiplied by its charge.
The averages, maxima, and minima of the outputs are processed with a final network, the event network, which outputs one variable, \qrgflat, which is in $[-1,1]$, with $\qtag = \sign(\qrgflat)$ and $r = \abs*{\qrgflat}$. 

We train \gflat using simulated events in which \PBtag decays generically according to known~\prdcom{\cite{PDG}} (if known) or assumed (otherwise) branching fractions and \PBsig decays to $\Pnu\APnu$, so that all reconstructed tracks and ECL clusters form the ROE.
The training data set consists of \num{5e+6} events; the independent validation data set consists of \num{8e+5} events.
We minimize binary cross-entropy loss with the Adam optimizer~\cite{Adam} and train with a one-cycle learning schedule~\cite{oneCLR}.

\begin{figure}[tb]
    \centering
    \includegraphics[width=\linewidth]{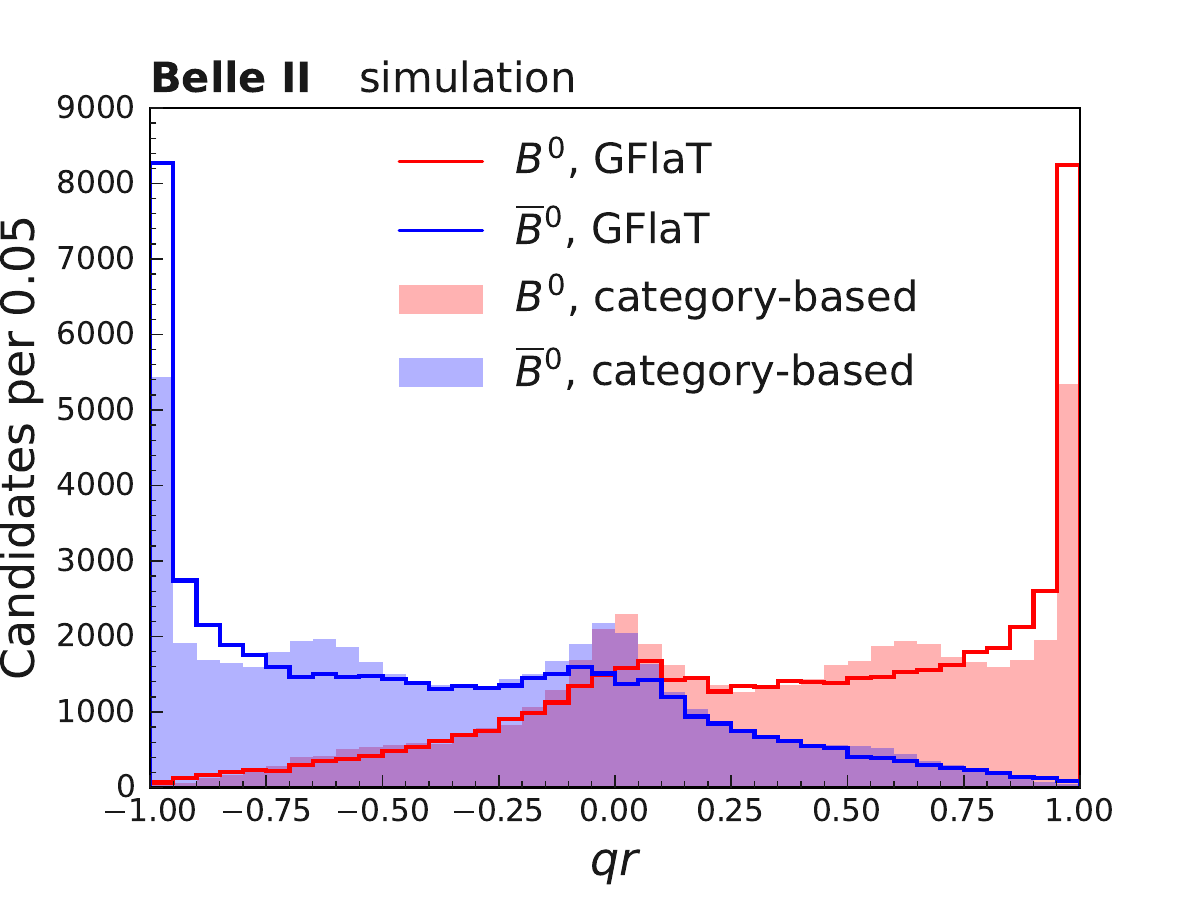}
    \caption{Distributions of $qr$ for true \PBzero and \APBzero from \gflat and the category-based flavor tagger in simulated data.}
    
    \label{fig:qr_comparison}
\end{figure}

Figure \ref{fig:qr_comparison} shows the $qr$ distributions for true \PBzero and \APBzero for independent test data consisting of \num{1e+5} events from \gflat and the category-based flavor tagger. 
The latter has more reliable tagging information than reported in Ref.~\cite{Belle-II:2021zvj}, due to recent improvements in particle identification and parameter tuning.
\gflat better distinguishes between \PBzero and \APBzero than the category-based flavor tagger: 
the peaks at $|qr|\approx1$ are higher and the bumps at $|qr|\approx0$ and $|qr|\approx0.65$ are smaller.

\begin{figure*}[tb]
    \centering
    \includegraphics[width=0.8\linewidth]{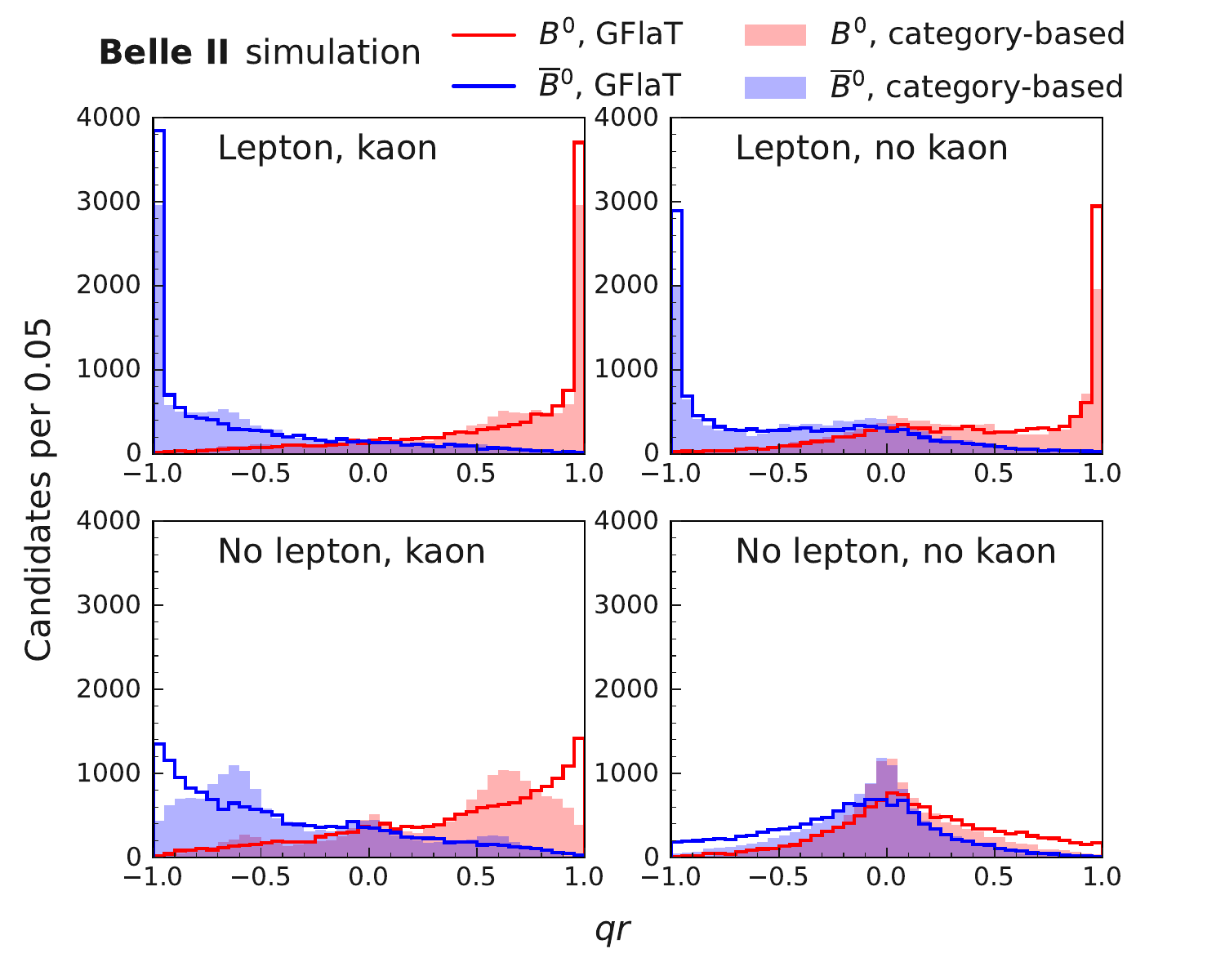}
    \caption{Distributions of $qr$ for true \PBzero and \APBzero from \gflat and the category-based flavor tagger for events classified according to the presence of charged leptons or charged kaons in the ROE in simulation data.
    }
    \label{fig:qr_comparison_category}
\end{figure*}

Figure \ref{fig:qr_comparison_category} shows the $qr$ distributions 
for events classified according to the presence of charged leptons or kaons in the ROE.
The ROE contains a charged lepton and a charged kaon in 22.2\% of events, a charged lepton and no charged kaon in 22.9\%, a charged kaon and no charged lepton in 31.5\%, and neither in 23.4\%.
The distributions indicate that performance is optimal when both a lepton and a kaon are present, with the contribution from leptons being particularly significant.
The distributions also reveal that the bump at $|qr|\approx0.65$ in the category-based flavor tagger is due to events with charged kaons, which indicates that flavor assignment in such events is less reliable since a 
\PKminus, predominantly associated with \APBzero decays, can also originate from a \PBzero decay, for example through decay to a \PDminus with $\PDminus\to\APKzero\PKminus$.
Since \gflat accounts for the relationships between final-state particles, it can better discern the origin of the tracks; and so its output does not peak at $|qr|\approx0.65$ for those events, but instead at $|qr|\approx1$.
Both flavor taggers perform poorly for events with neither a charged lepton nor a charged kaon, consisting mostly of pions, but \gflat's output still exhibits a visible improvement.
A charged pion from \PBzero decay, such as $\PBzero\to\PDminus\Ppiplus$,
or through an intermediate resonance that decays via the strong force, correlates with the \PB flavor. 
The \gflat algorithm exploits this correlation more effectively to improve performance.

\section{Calibration and performance}
\label{sec:CalibrationAndPerformance}
We evaluate \gflat's performance using events in which \PBsig decays to the $\PDoptstarminus\Ppiplus$ final state.
The flavor of \PBsig is determined by the charge of the pion, neglecting the wrong-sign contribution.
We fit the $\Delta t$ probability density model to the background-subtracted $\Delta t$ distribution, accounting for resolution effects, to determine the wrong-tag probability $w$ in each $r$ bin.
We subtract the background with \sWeight~\cite{Pivk:2004ty, Dembinski:2021kim} using the \PB energy as a discriminating variable.

We reconstruct \PDminus candidates via $\PDminus\to\PKplus\Ppiminus\Ppiminus$ and \PDstarminus via $\PDstarminus\to\APDzero\Ppiminus$ with $\APDzero\to\PKplus\Ppiminus$, $\PKplus\Ppiminus\Ppizero$, or $\PKplus\Ppiminus\Ppiplus\Ppiminus$.
Tracks 
must originate from the IR and 
have polar angles within the CDC.

We reconstruct \Ppizero candidates via $\Ppizero\to\Pphoton\Pphoton$, forming photon candidates from ECL clusters not associated with any tracks.
To suppress beam-background photons, we require each cluster have an energy greater than \SI{120}{MeV}, \SI{30}{MeV}, or \SI{80}{MeV} if it is in the forward, barrel, or backward region of the ECL,
which corresponds to the lab-frame polar angle ranges 
\SIrange{12.4}{31.4}{\degree}, \SIrange{32.2}{128.7}{\degree}, and \SIrange{130.7}{155.1}{\degree}, respectively.
The angle between the photon momenta must be less than \SI{52}{\degree} in the lab frame and the diphoton mass must be in the range \SIrange{121}{142}{MeV}, which is centered on the known \Ppizero mass and is six units of diphoton mass resolution wide.

One of the \PD's decay products must be consistent with being a \PKplus, but no particle-identification requirements are placed on the other charged particles.
Each \PDminus candidate must have a mass in \SIrange{1.860}{1.880}{GeV}, which is centered on the known \PDminus mass and is a $\pm 3\sigma$ range, 
with $\sigma$ being the mass resolution.
Each \APDzero candidate reconstructed from $\PKplus\Ppiminus(\Ppiplus\Ppiminus)$ must have a mass in \SIrange{1.845}{1.885}{GeV}, which is centered on the known \APDzero mass and is a $\pm 5\sigma$ range.
Each \APDzero candidate reconstructed from $\PKplus\Ppiminus\Ppizero$ must have a mass in \SIrange{1.810}{1.895}{GeV}, which is an asymmetric range of $+2.5\sigma$ and $-4\sigma$ around the known \APDzero mass to account for energy losses in photon reconstruction.

The \Ppiminus from a \PDstarminus candidate decay must have momentum below \SI{300}{MeV} in the $\APelectron\Pelectron$ center-of-mass~(c.m.)~frame.
Each \PDstarminus candidate must have an energy release, $m(\PDstarminus) - m(\APDzero) - m_{\Ppiminus}$, in \SIrange{4.6}{7.0}{MeV}, which is centered around the known energy release and six units of its resolution wide.

We reconstruct a \PBzero candidate from a \PDoptstarminus candidate and a track that is consistent with being a \Ppiplus.
For each \PBzero candidate, we fit the trajectories and momenta of its decay products according to its decay chain with \softwarelibrary{TreeFit}~\cite{Krohn:2019dlq}, constraining the \PBzero to originate from the IR and the \PDoptstar to its known mass~\cite{PDG}.
We reject \PBzero candidates whose fits do not converge. The fraction of rejected signal candidates is 0.4\%.
We define the signal region from a beam-constrained mass
\begin{equation}
    M\Sub{bc} \equiv \sqrt{E\Sub{beam}^2 - \abs*{\vec{p}}^2}
\end{equation}
and energy difference,
$\Delta E \equiv E - E\Sub{beam}$,
where $E\Sub{beam}$, $E$, and $\vec{p}$ are the beam energy and \PBzero energy and momentum in the c.m.~frame, respectively.
The criteria for the signal region are $M\Sub{bc} > \SI{5.27}{GeV}$ and $\Delta E \in \SIrange{-0.10}{0.25}{GeV}$.

We determine the decay position of \PBtag by fitting the trajectories of ROE tracks with \softwarelibrary{Rave}~\cite{Waltenberger:2008zza}.
Unlike \softwarelibrary{TreeFit}, \softwarelibrary{Rave} accounts for the unknown \PBtag decay chain by reducing the impact of a displaced vertex due to potential intermediate \PD's, 
constraining the \PBtag vertex position to be consistent with the origin and direction of \PBsig.
We reject events in which this fit does not converge, which rejects 3.4\% of the signal events.

To suppress events not coming from $\APelectron\Pelectron\to\PB\APB$, such as $\APelectron\Pelectron\to\Pquark\APquark$, we exploit their topological differences, by requiring the ratio of the second to the zeroth Fox-Wolfram moment, $R_2$, be less than \num{0.4}~\cite{Fox:1978vu}. After applying all selection requirements, the average number of candidates per event is 1.05 and all candidates are retained. 

Events passing the above criteria are either correctly identified \PBsig decays or 
backgrounds from $\PB\APB$ and $\Pquark\APquark$ events.
To separate signal from background, we fit to the $\Delta E$ distribution using an extended unbinned likelihood, combining data from \PBsig and \APBsig and all $r$ intervals.

We model the signal contribution as the sum of a Gaussian function and a double-sided Crystal-Ball function~\cite{Skwarnicki:1986xj}.
Their parameters and their admixture are fixed to values obtained from fitting to simulated data, but a common shift of their peak values and common scaling of their widths are left free 
to account for differences between data and simulation.

Events in which \PBsig decays to the $\PDoptstarminus\PKplus$ final state, with the \PKplus misidentified as a \Ppiplus, peak at \SI{-50}{MeV} in the $\Delta E$ distribution.
According to simulation studies, the fraction of these events to the signal is $2.5\%$.
We model this contribution as a double-sided Crystal Ball function, whose parameters are fixed to values obtained from fitting to simulated data, including the ratio of its yield to the signal, except for the shift of its peak value and the scaling of its width, which are the same as for the signal.
Since these events have the same $\Delta t$ distribution as $\PBzero\to\PDoptstarminus\Ppiplus$, we use this contribution as signal in the \sWeight calculation.

We model the $\PB\APB$ background contribution as a second-order polynomial, with the ratio of its yield to that of the signal fixed to a value obtained from simulated data.
We model the $\Pquark\APquark$ background contribution as an exponential function.
To constrain the parameters of the $\Pquark\APquark$ component, we simultaneously fit to the $\Delta E$ distribution in a sideband, $M\Sub{bc} \in \SIrange{5.20}{5.24}{GeV}$, populated predominantly by $\Pquark\APquark$ events.
We confirm via simulation studies that the $\Delta E$ distributions of the $\Pquark\APquark$ component in the signal and sideband regions 
are sufficiently similar to warrant a simultaneous fit.

\begin{figure}[tb]
  \centering
  \includegraphics[width=\linewidth]{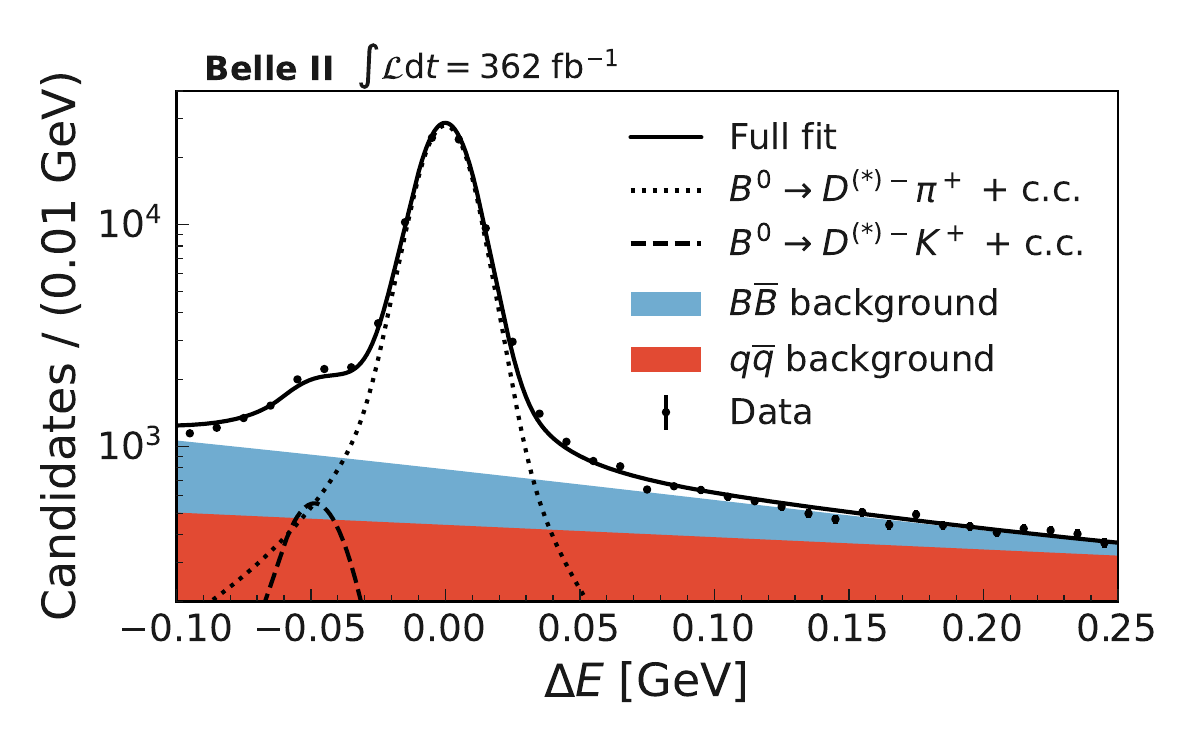}%
  \hfill%
  \includegraphics[width=\linewidth]{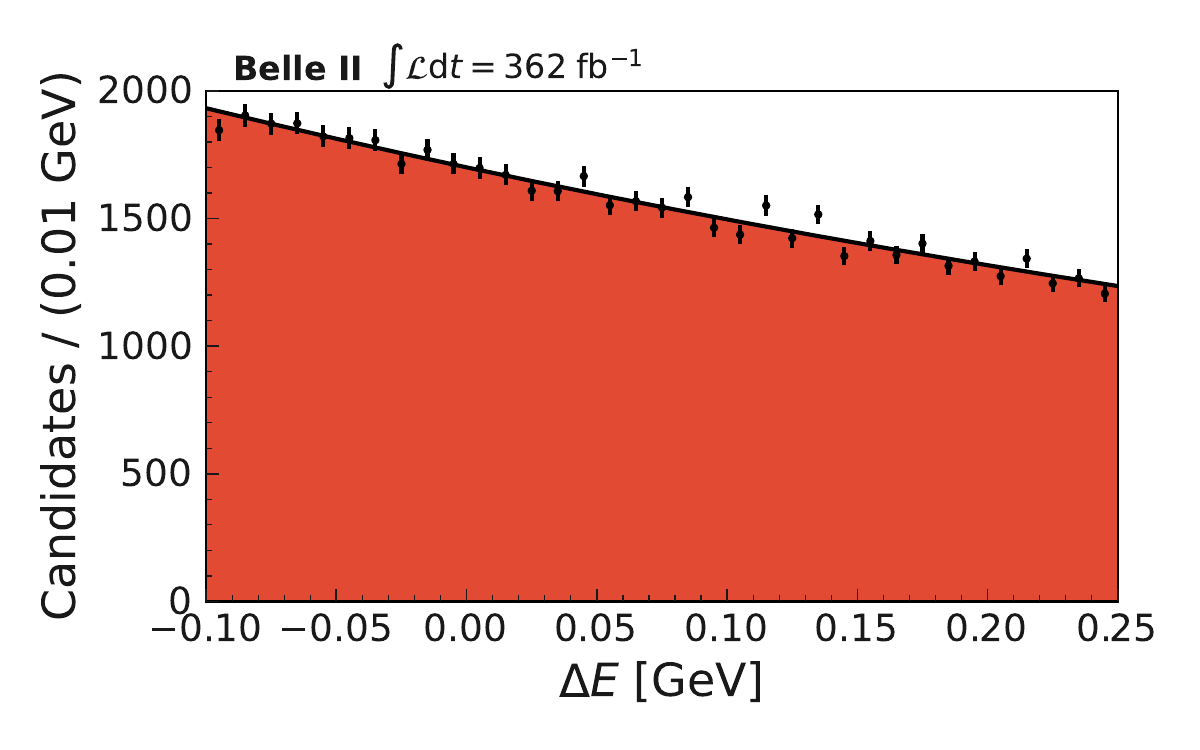}
  \caption{Distributions of $\Delta E$ for $\PBzero\to\PDoptstarminus\Ppiplus$ reconstructed in data in the signal region~(top) and sideband~(bottom) and the best-fit function, with background components stacked.}
  \label{fig:dE_fitFullSB_data}
\end{figure}

Figure~\ref{fig:dE_fitFullSB_data} shows the $\Delta E$ distributions in the signal region and sideband and the fit results.
The fit agrees well with the data.
Yields in the signal region are \num{77130 \pm 320} events for the signal (for the sum of the $\PDoptstarminus\Ppiplus$ and $\PDoptstarminus\PKplus$ final states), \num{8620 \pm 40} for the $\PB\APB$ background, and \num{14200 \pm 230} for the $\Pquark\APquark$ background.

We modify equation~(\ref{eq:dt_flav_perf}) to account for 
differences in the wrong-tag probabilities for \PBtag and \APBtag, by introducing $w(\PBzero) \equiv \wavg + \tfrac12\Delta w$ and $w(\APBzero) \equiv \wavg - \tfrac12 \Delta w$ and reconstruction efficiency asymmetries for \PBsig and \PBtag, \asig and \atag,
with $a_{x} \equiv [\varepsilon(\PBzerox) - \varepsilon(\APBzerox)] / [\varepsilon(\PBzerox) + \varepsilon(\APBzerox)]$, where $x$ indicates `tag' or `sig',
\begin{alignat}{1}
    &P(\Delta t,\qsig,\qtag) = \nonumber\\
    &~(1 + \asig \qsig) \frac{e^{-\abs*{\Delta t} / \tau}}{4\tau} \big\{1 + \qtag [\atag (1 - 2\wavg) - \Delta w] \nonumber\\
    &~+ \qsig \qtag \qty(1 - 2\wavg + \qtag\atag - \atag\Delta w) \cos(\Delta m_{\Pdown} \Delta t) \big\}.
    \label{eq:MixingMain}
\end{alignat}

We determine \asig by fitting the $\Delta E$ distributions for \PBsig and \APBsig separately, using the same model as for their combined fit, without selection criteria on \PBtag to avoid a bias from using \PBtag information.
We measure 
\asig = \SI{-2.53 \pm 0.39}{\percent},
which we attribute to charge asymmetries in kaon identification and low-momentum track finding.

We calculate a per-candidate signal probability using \sWeight from the $\Delta E$-fit results, allowing us to statistically subtract background contributions to the $\Delta t$ distributions.
This requires that $\Delta E$, $\Delta t$, and $r$ be independent, 
which is confirmed in simulation studies.

We calculate $\Delta t$ from the distance, $\Delta \ell$, of the \PBsig vertex from that of \PBtag along the \FourS boost direction,
\begin{equation}
    \Delta t = \frac{\Delta\ell}{\beta\gamma\, \gamma_{\PB}},
\end{equation}
where $\beta\gamma=0.28$ is the Lorentz boost of the \FourS in the lab frame and $\gamma_{\PB}=1.002$ is the Lorentz factor of the \PB in the c.m.~frame.

To account for resolution and bias in measuring $\Delta\ell$, we convolve equation~(\ref{eq:MixingMain}) with the resolution function introduced in Ref.~\cite{hadpaper}.
The resolution function consists of a core component modeled by a Gaussian function, a tail component modeled by a weighted sum of a Gaussian and two exponentially modified Gaussian functions, and an outlier component modeled by a Gaussian function.
Parameters of the resolution function are shared by all $r$ bins, 
except for the highest $r$ bin. This bin is mostly populated by semileptonic \PBtag decays, which have a better resolution.

\begin{figure*}
    \centering
    \includegraphics[width=0.385\linewidth]{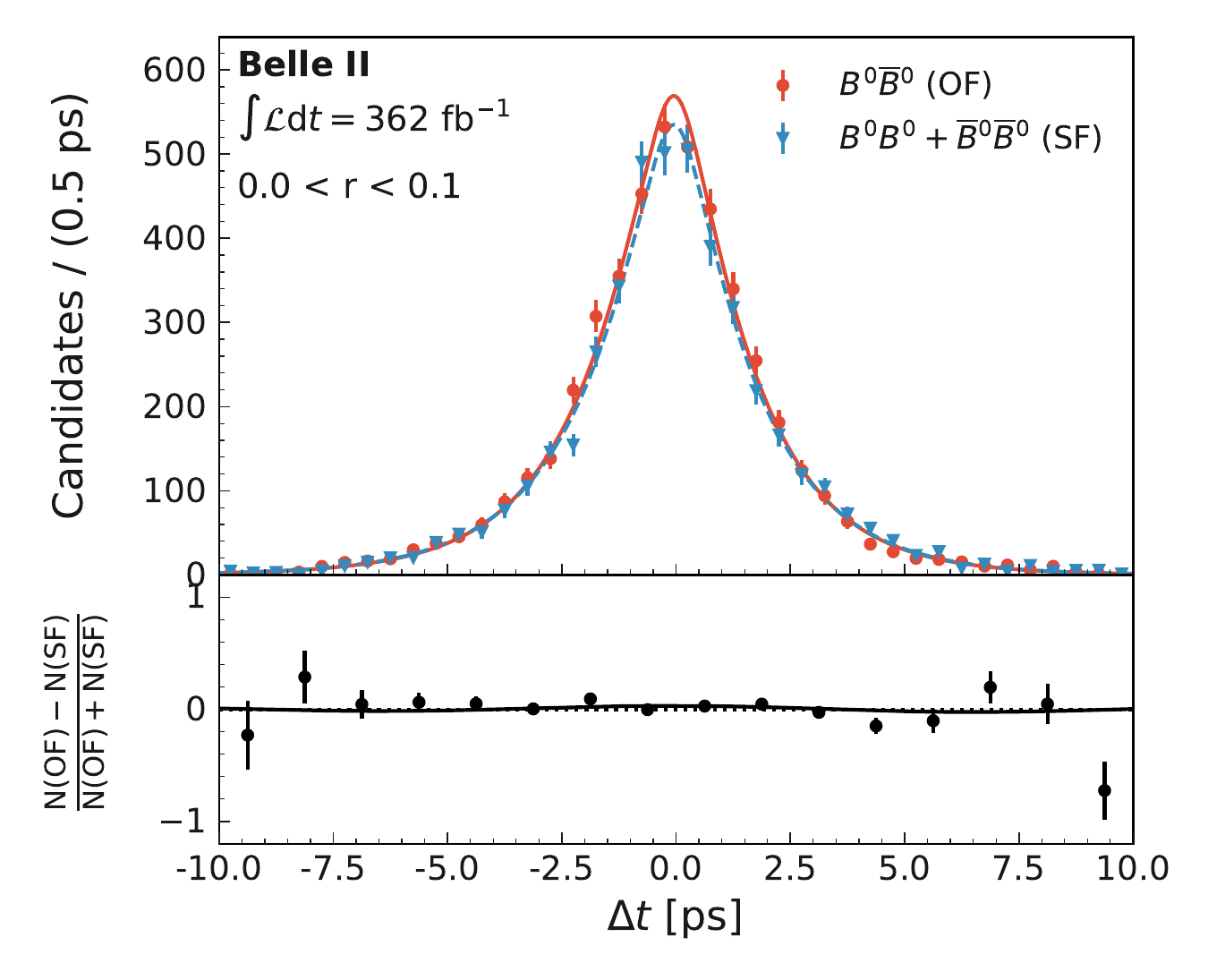}
    \includegraphics[width=0.385\linewidth]{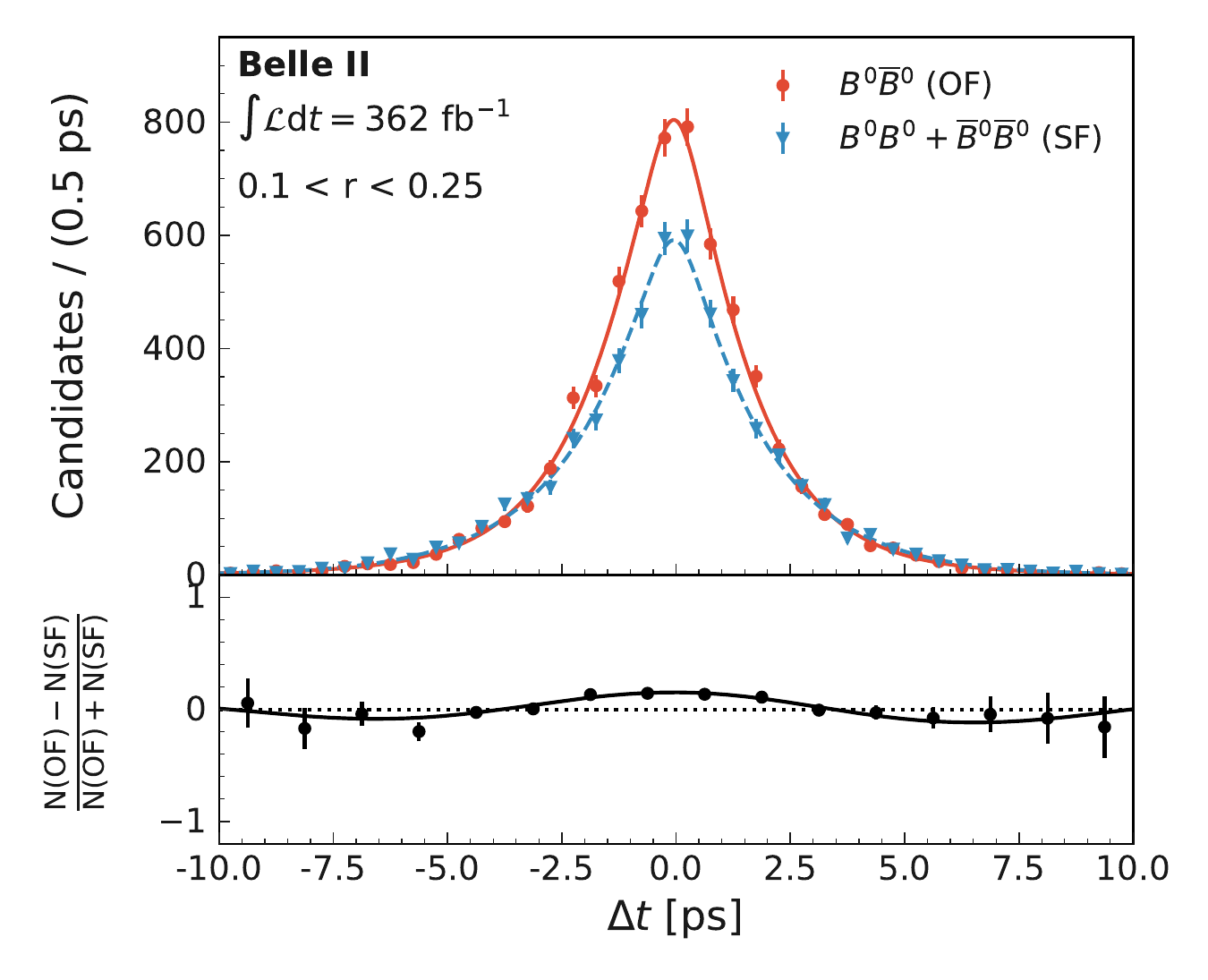}
    \includegraphics[width=0.385\linewidth]{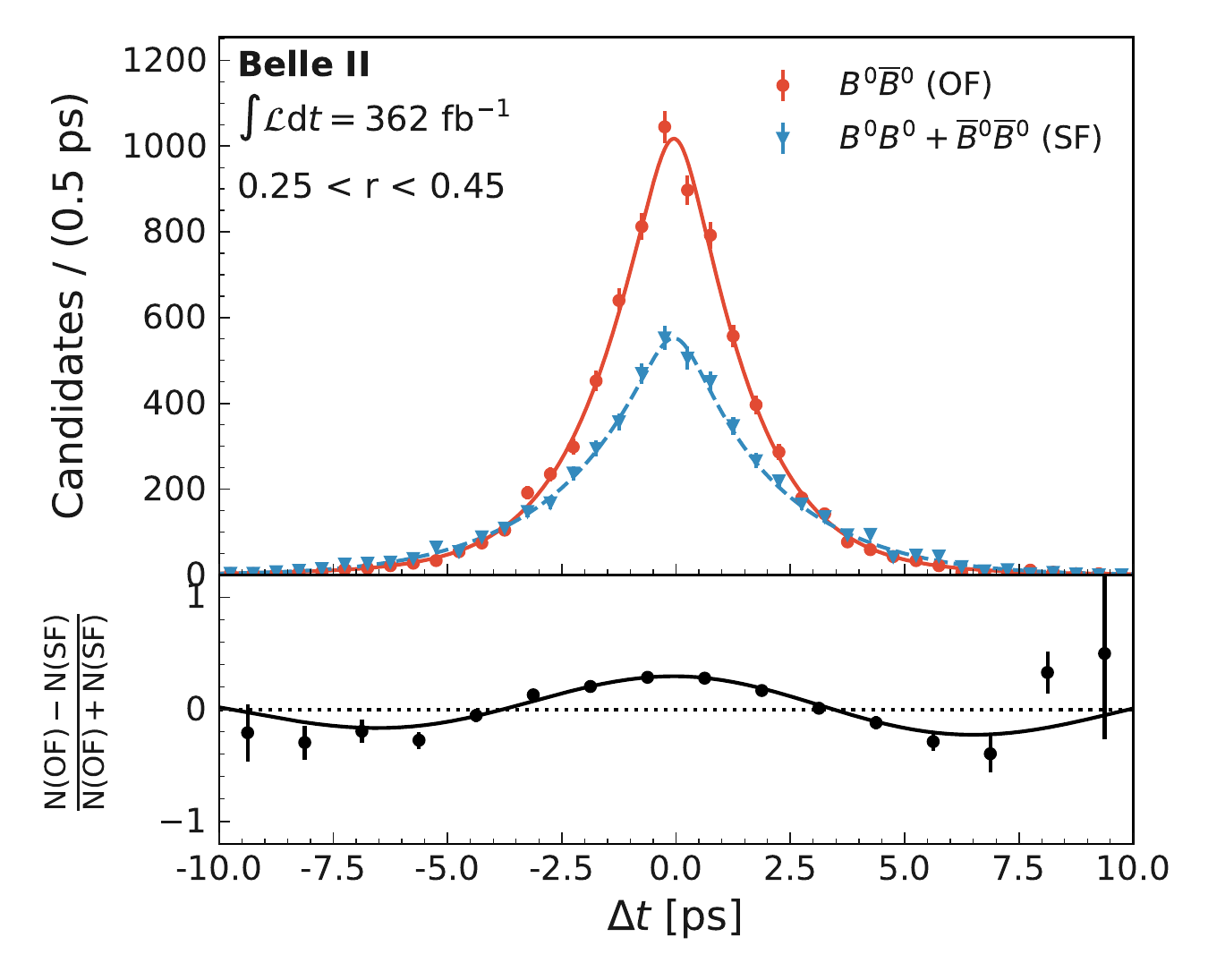}
    \includegraphics[width=0.385\linewidth]{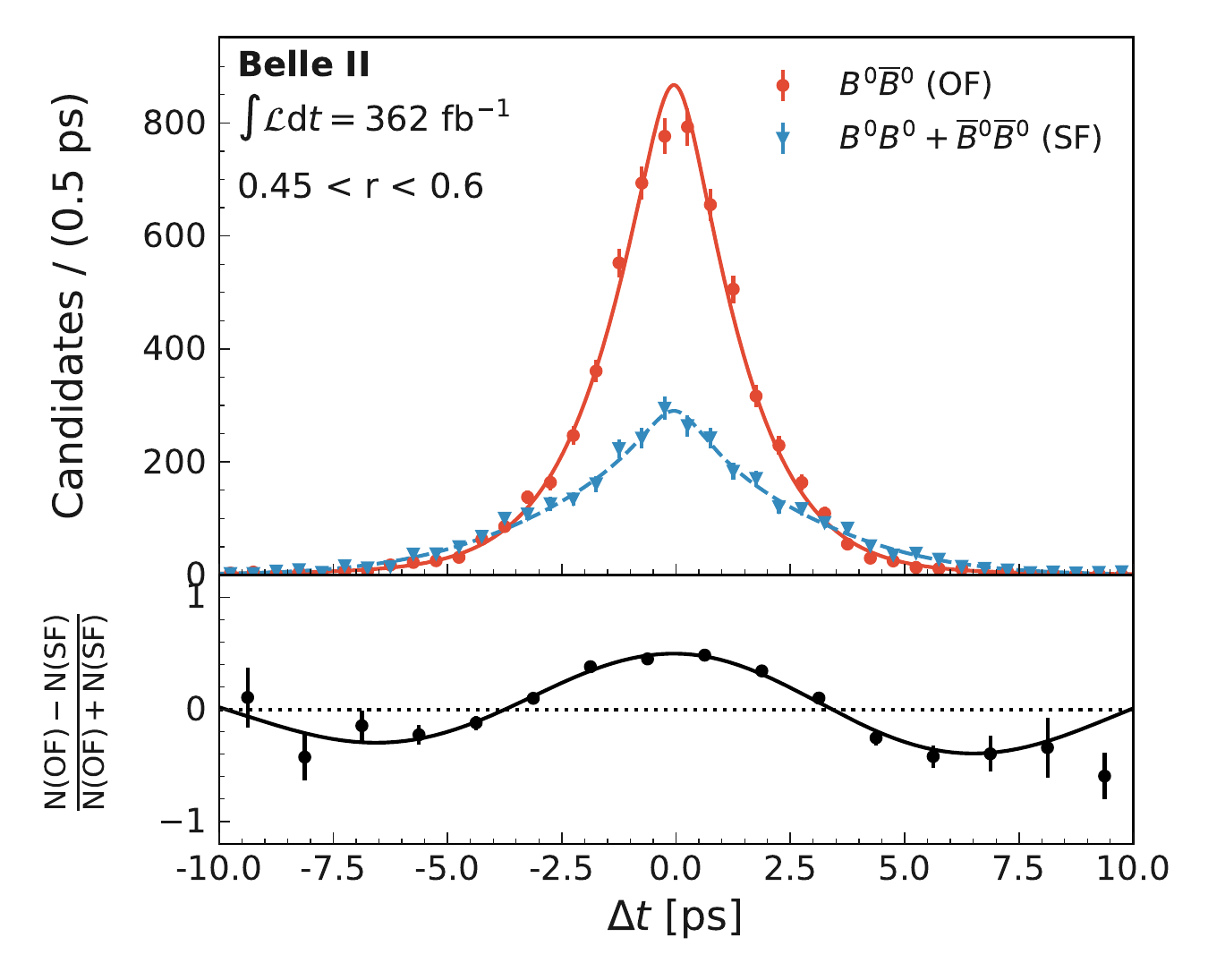}
    \includegraphics[width=0.385\linewidth]{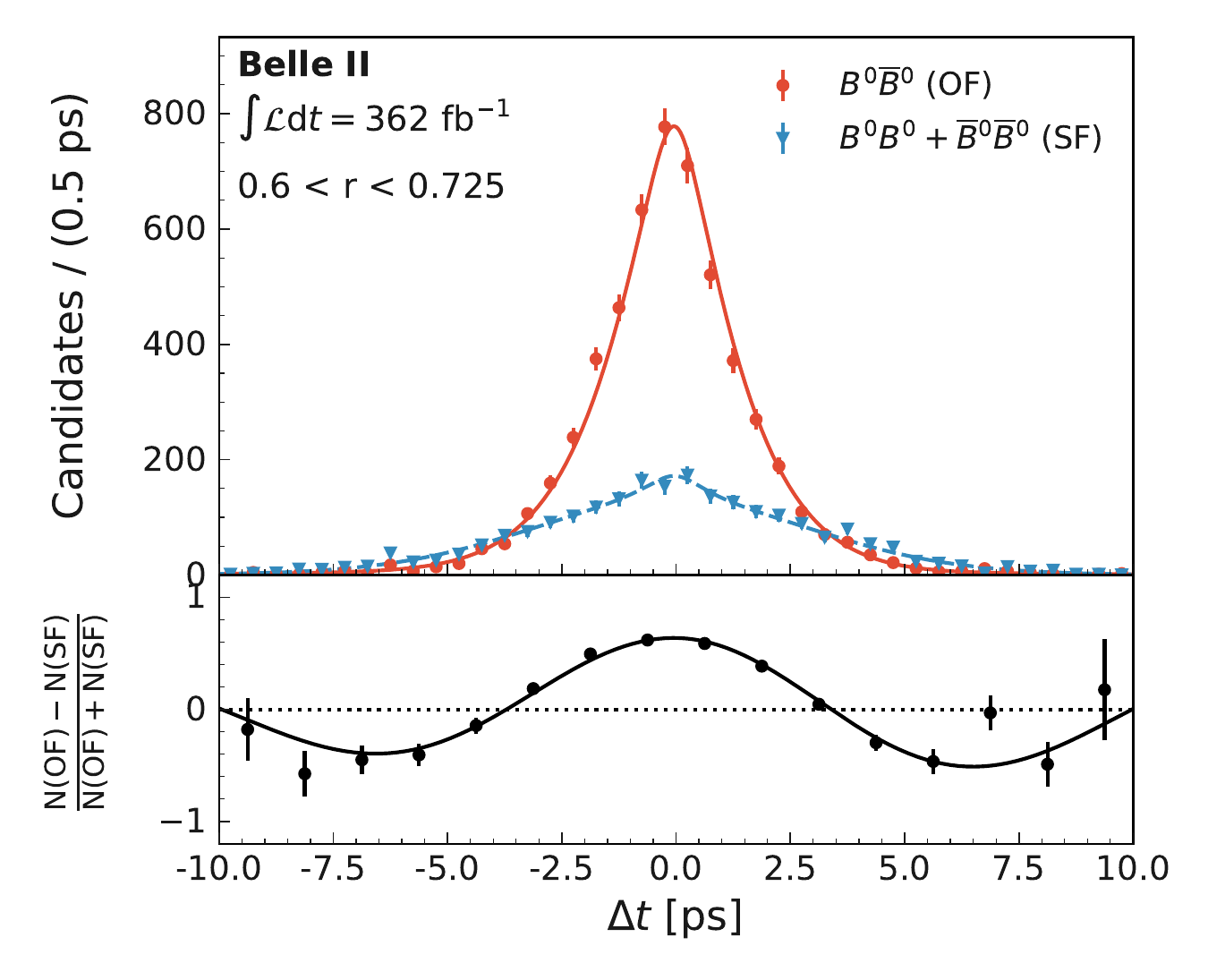}
    \includegraphics[width=0.385\linewidth]{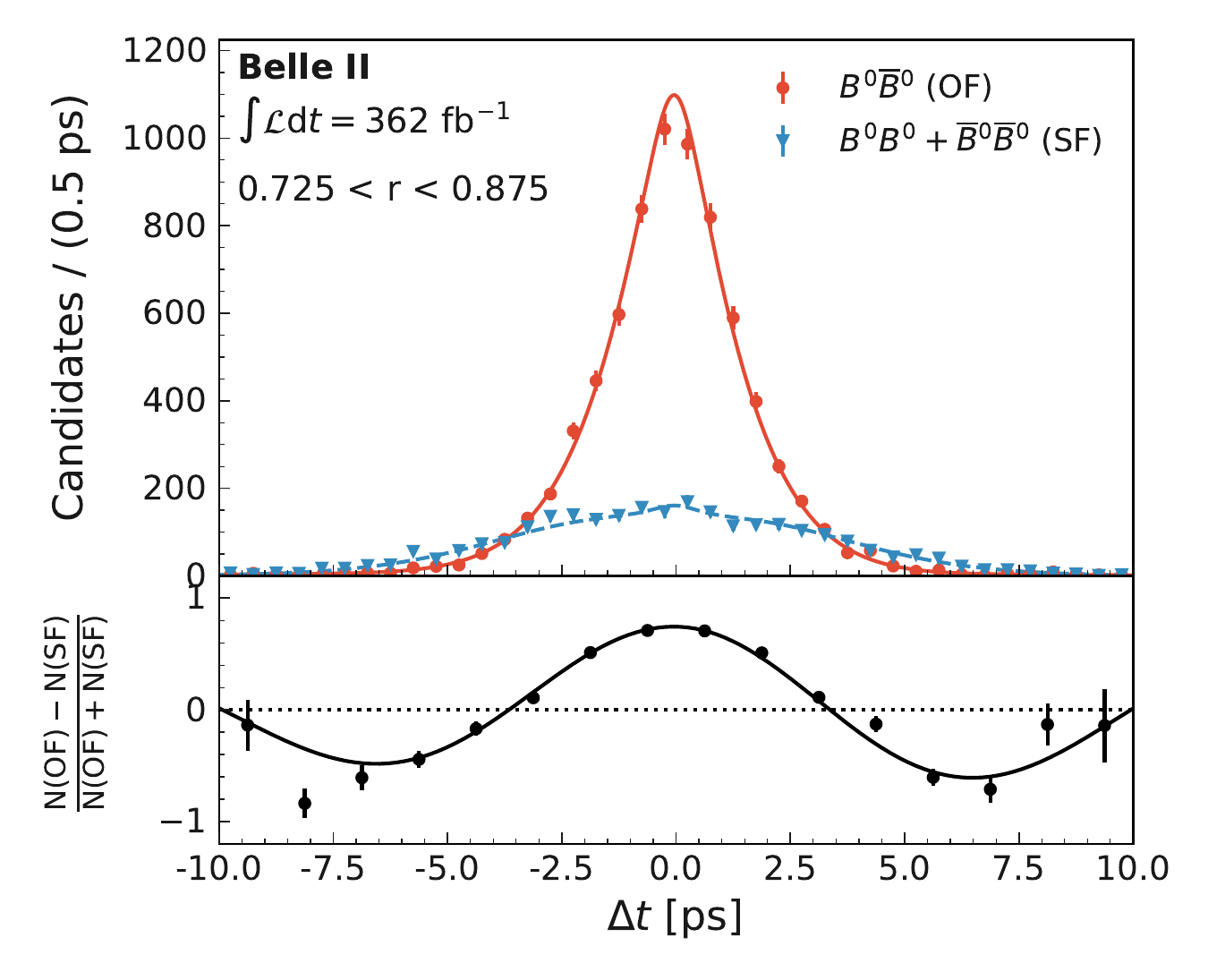}
    \includegraphics[width=0.385\linewidth]{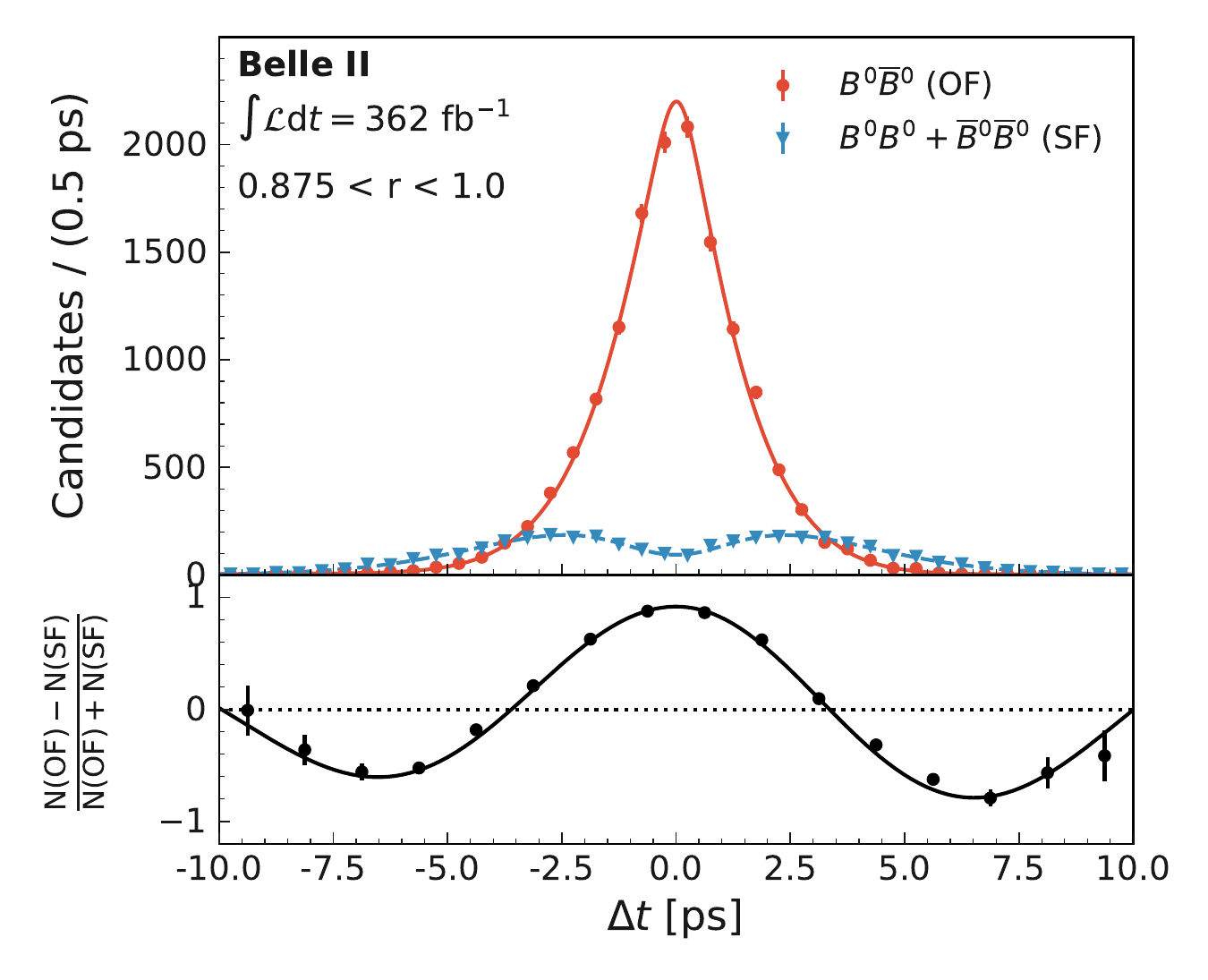}
    \caption{Background-subtracted $\Delta t$ distributions of $\PBzero\to\PDoptstarminus\Ppiplus$ reconstructed in data in each of the seven $r$ intervals~(points) and the best-fit functions~(lines) for opposite- and like-flavor \PB pairs with the corresponding asymmetries.}
    \label{fig:calibration_mixing}
\end{figure*}

We fit simultaneously to the binned background-subtracted $\Delta t$ distributions in 28 subsets of the data defined by the 7 $r$ intervals, 2 flavors of \PBsig, and 2 flavors of \PBtag.
The fit has seven free resolution-function parameters and 21 free flavor-tagger parameters, \atag, \wavg, and $\Delta w$ in each of the 7 $r$ bins.
The uncertainty on the $\Delta t$ measurement, $\sigma_{\Delta t}$, is computed for each event and is a conditional variable in the resolution function. 
We use a histogram with 500 bins in each data subset as the probability density function for this variable.
We fix $\Delta m_{\Pdown}$ and $\tau$
to their world average values~\cite{PDG}.
Figure~\ref{fig:calibration_mixing} shows the $\Delta t$ distribution in each $r$ interval and the result of the fit.

\begin{figure}[t]
    \centering
    \includegraphics[width=\linewidth]{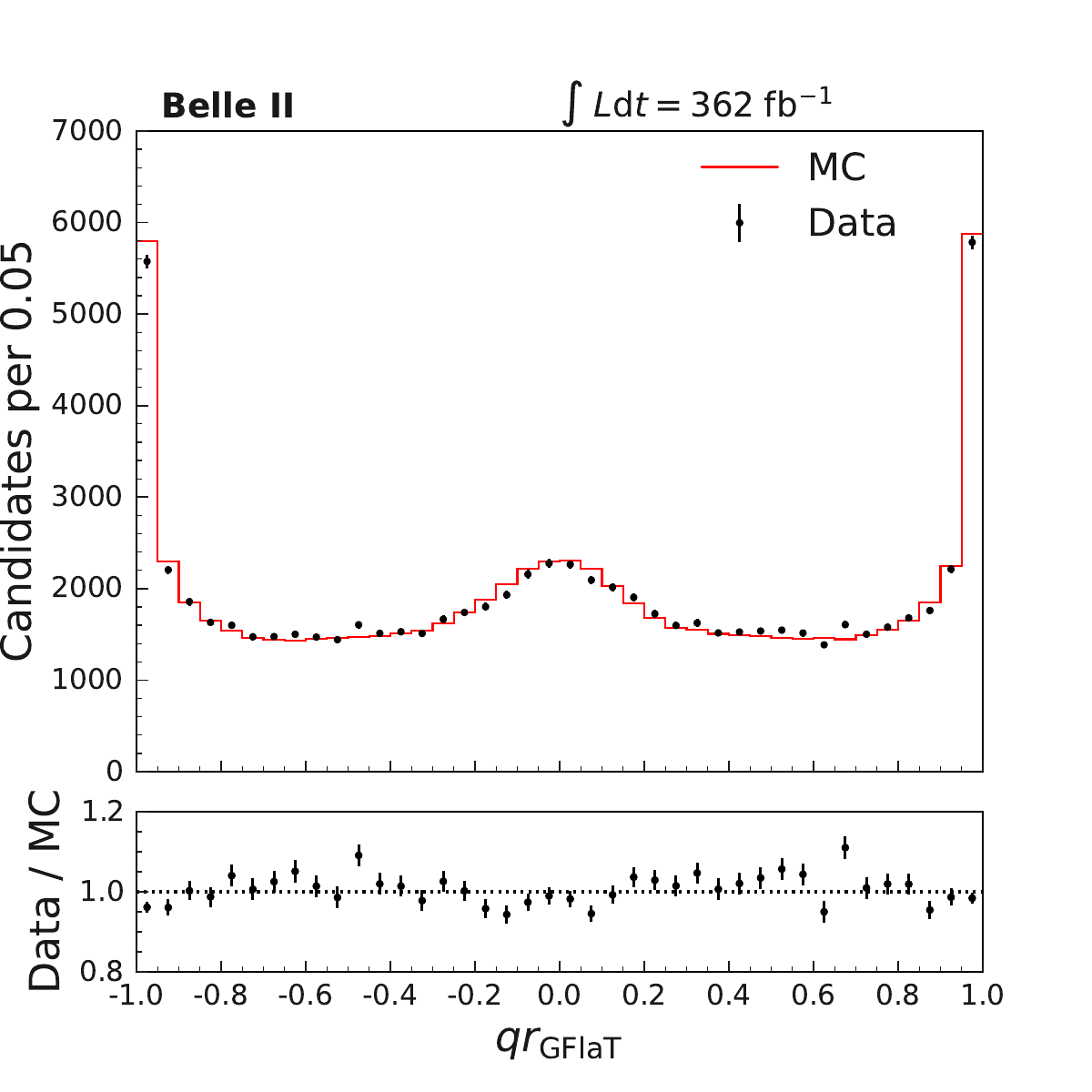}
    \caption{Distributions of \qrgflat for $\PBzero\to\PDoptstarminus\Ppiplus$ in background-subtracted data and correctly reconstructed simulated events normalized to the data signal yield.}
    \label{fig:qr_data_MC_362}
\end{figure}

\begin{figure}[htbp]
    \centering
    \includegraphics[width=\linewidth]{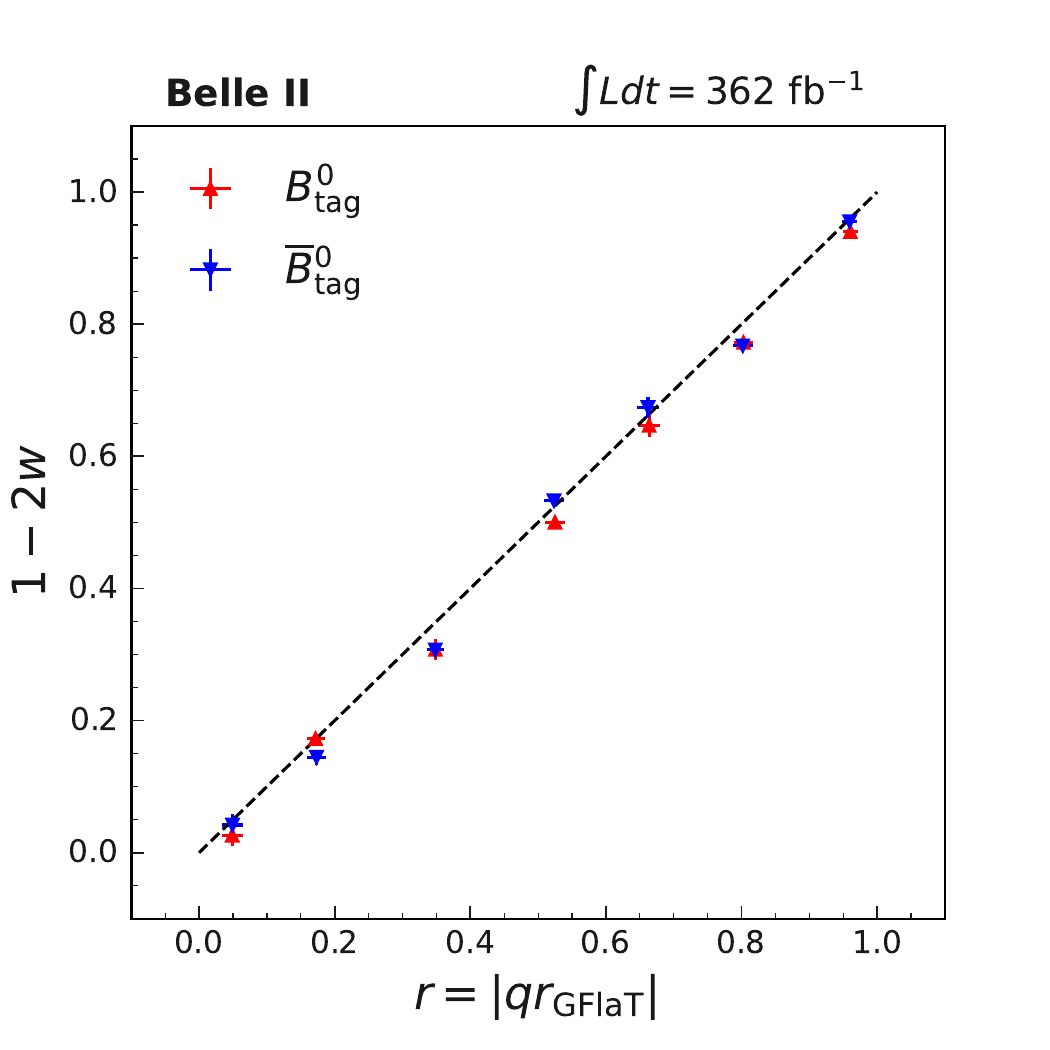}
    \caption{Dilution factors $1 - 2w$ of $\PBzero\to\PDoptstarminus\Ppiplus$ as functions of their \gflat predictions, $r$ for \PBzerotag, $1 - 2\wavg - \Delta w$, and \APBzerotag, $1 - 2\wavg + \Delta w$; the dashed line shows $r = 1 - 2 w$.}
    \label{fig:dilution}
\end{figure}

Figure~\ref{fig:qr_data_MC_362} shows the \qrgflat distribution in background-subtracted data and correctly reconstructed simulated events normalized to the data signal yield.
Figure~\ref{fig:dilution} shows the dilution factors, $1 - 2w_i$, for each $r$ bin $i$ for both \PBzerotag and \APBzerotag.
It shows that $r$ is a good estimator of $1-2w$ for both tag flavors.
The effective tagging efficiency is $\efftag = (\num{37.40 \pm 0.43})\%$, where the uncertainty is statistical only.
Table~\ref{tab:data_fitted_parameters} in Appendix \ref{sec:app_params} lists \atag, \wavg, and $\Delta w$ for each $r$ bin. 


%%%%%%%%%%%%%%%%%%%%%%%%%%%%%%%%%%
\section{\texorpdfstring{Measurement of $\sin2\phi_1$ in $\PBzero\to\PJpsi\PKshortzero$}{sin2phi1}}
\label{sec:MeasurementOfSin2phi1}

We demonstrate \gflat by measuring \SCP and  \CCP in $\PBzero\to\PJpsi\PKshortzero$ decays.
We reconstruct \PJpsi candidates via $\PJpsi\to\APelectron\Pelectron$ or $\APmuon\Pmuon$.
The leptons must fulfill the same track requirements as described for the decay products of $\PBzero\to\PDoptstarminus\Ppiplus$ and be consistent with both being electrons or both being muons.
To account for energy loss due to bremsstrahlung, the four-momenta of photons with lab-frame energy in \SIrange{75}{1000}{MeV} detected within \SI{50}{mrad} of the initial direction of an electron are added to the electron's four-momentum.
Each $\PJpsi\to\APelectron\Pelectron$ candidate must have a mass in \SIrange{2.90}{3.14}{GeV}; each $\PJpsi\to\APmuon\Pmuon$ candidate must have a mass in \SIrange{3.00}{3.14}{GeV}. 
The resolutions at masses above and below the known $\PJpsi$ mass are \SI{8.0}{MeV} and \SI{9.0}{MeV} for electron pairs and \SI{6.3}{MeV} and \SI{8.3}{MeV} for muon pairs.

We reconstruct \PKshortzero candidates via $\PKshortzero\to\Ppiplus\Ppiminus$.
The pions must have polar angles within the CDC.
Each \PKshortzero candidate must have a mass in the range \SIrange{0.45}{0.55}{GeV}, a successful vertex fit, and a decay vertex displaced from the IR by at least five units of the displacement's uncertainty. The reconstructed $\PKshortzero$ mass resolution is \SI{2.0}{MeV}.
\prdcom{Possible bias related to \CP violation in the $\PKzero$-$\APKzero$ system, as well as kaon regeneration are expected to be very small and are neglected in this analysis~\cite{Bjorn:2019kov}.}

We fit the trajectories and momenta of \PBzero decay products with \softwarelibrary{TreeFit}, constraining the \PBzero to originate from the IR and the \PJpsi to have its known mass~\cite{PDG}.
Each \PBzero candidate must have $M\Sub{bc}$ greater than \SI{5.27}{GeV} and $\Delta E$ in \SIrange{-0.10}{0.25}{GeV}.
The \PBtag vertex position is determined as described for $\PBzerosig\to\PDoptstarminus\Ppiplus$ above.
We require $R_2$ be less than 0.4 to remove $\Pquark\APquark$ background. After applying all selection requirements, the average number of candidates per event is 1.01. All candidates are retained for further analysis.

To validate our analysis, we also measure \SCP and \CCP for $\PBzero\to\PJpsi\PKstar(892)^0$.
\prdcom{We expect $\SCP = 0$, as this decay mode is flavor-specific, and $\CCP = 0$ --- as defined in Eq.~\ref{eq:dt_cp_perf} ---  as with $\PBzero\to\PJpsi\PKshortzero$.}
Hereafter, $\PKstar(892)^0$ is written as $\PKstarzero$.
We reconstruct \PKstarzero candidates via $\PKstarzero\to\PKplus\Ppiminus$, requiring the positively charged particle be consistent with a \PKplus and the negatively charged particle be consistent with a \Ppiminus.
Each \PKstarzero candidate must have a mass in \SIrange{0.8}{1.0}{GeV}, 
corresponding to approximately four times the \PKstarzero natural width~\cite{PDG}. 
All selection criteria on \PJpsi and \PBzero candidates are the same as for $\PBzero\to\PJpsi\PKshortzero$, except that the \PBzero must have $\Delta E$ in a reduced range, \SIrange{-0.10}{0.10}{GeV}, to reject background from $\PBplus\to\PJpsi\PKplus$ with a \Ppiminus from \PBtag reconstructed as part of \PBsig.

We perform extended unbinned likelihood fits to the $\Delta E$ distributions
to determine signal and background yields and shapes that we use to statistically isolate the signal $\Delta t$ distributions using \sWeight.
We model the signal components as double-sided Crystal-Ball functions with tail parameters fixed to values determined from fits to simulated data and peak values and widths 
freely determined by the fits to data.
We model the background components
taking into account both $\PB\APB$ and $\Pquark\APquark$,
as exponential functions, whose parameters are freely determined by the fits to data.

\begin{figure}[t]
  \centering
  \includegraphics[width=\linewidth]{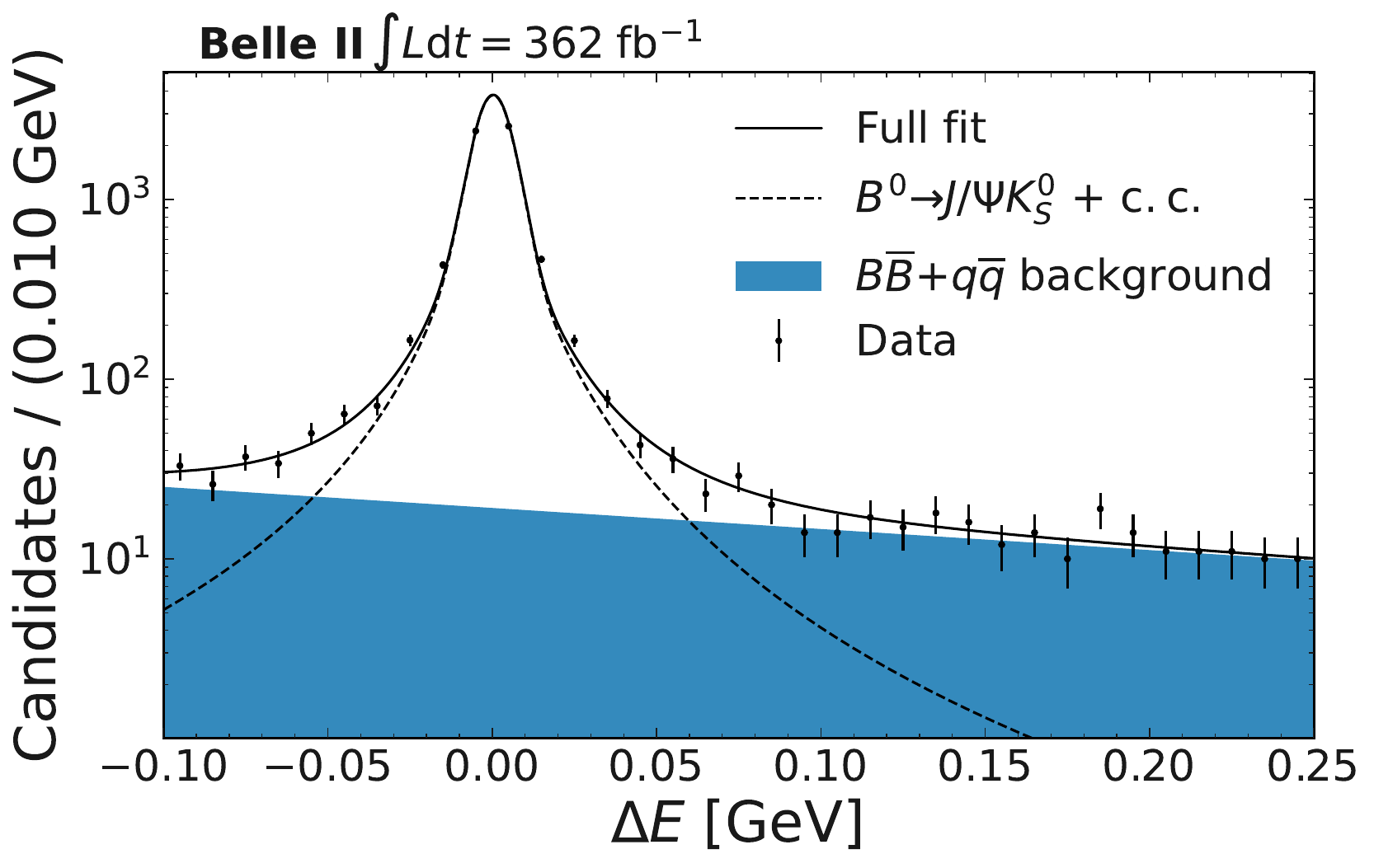}%
  \hfill%
  \includegraphics[width=\linewidth]{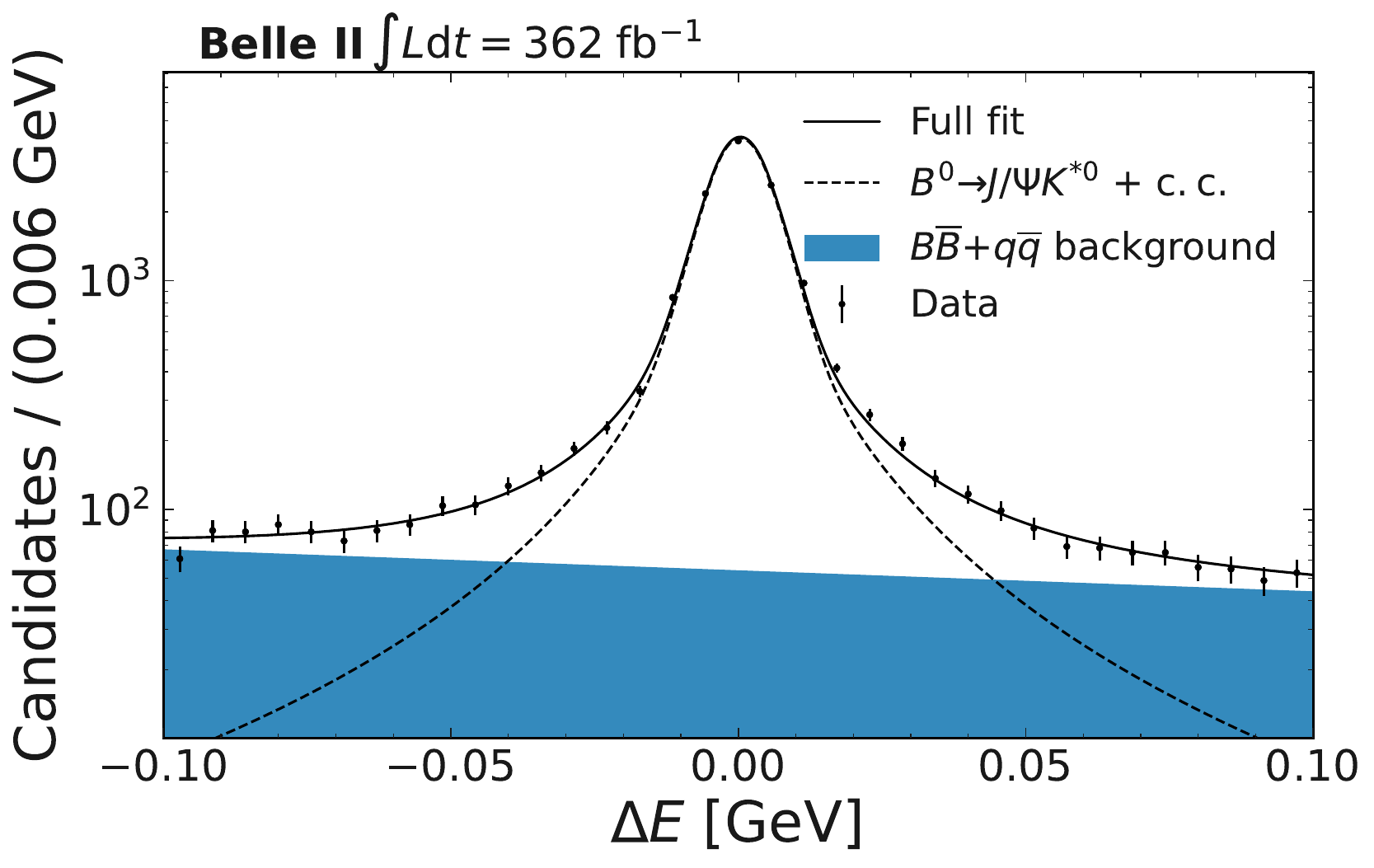}
  \caption{Distributions of $\Delta E$ for $\PBzero\to\PJpsi\PKshortzero$~(top) and $\PBzero\to\PJpsi\PKstarzero$~(bottom) and the best-fit functions.}
  \label{fig:defit_realdata}
\end{figure}

Figure~\ref{fig:defit_realdata} shows the $\Delta E$ distributions and the fit results.
The best-fit results agree well with the data.
For $\PBzero\to\PJpsi\PKshortzero$, the signal yield is \num{6390 \pm 90} and the background yield is \num{570 \pm 40}.
For $\PBzero\to\PJpsi\PKstarzero$, the signal yield is \num{12660 \pm 130} and the background yield is \num{1900 \pm 70}.

We determine \SCP and \CCP by performing a simultaneous fit to the background-subtracted $\Delta t$ distributions in 14 subsets defined by the 7 $r$ intervals and 2 flavors of \PBtag.
To take into account detection and tagging asymmetries, we modify equation~(\ref{eq:dt_cp_perf}),
\begin{alignat}{1}
    P(\Delta t,\qtag) = {} &\frac{e^{-\abs*{\Delta t}/\tau}}{4\tau}  \Big\{
        1 + \qtag \qty[\atag (1 - 2\wavg) - \Delta w] \nonumber\\
    &~+ \qtag \qty(1 - 2\wavg + \qtag \atag - \atag \Delta w) \nonumber\\
    &~ \times \qty[\SCP\sin(\Delta m_{\Pdown} \Delta t) - \CCP\cos(\Delta m_{\Pdown} \Delta t)] \Big\}.
    \label{eq:fobs}
\end{alignat}

\begin{figure}[t]
  \centering

  \includegraphics[width=\linewidth]{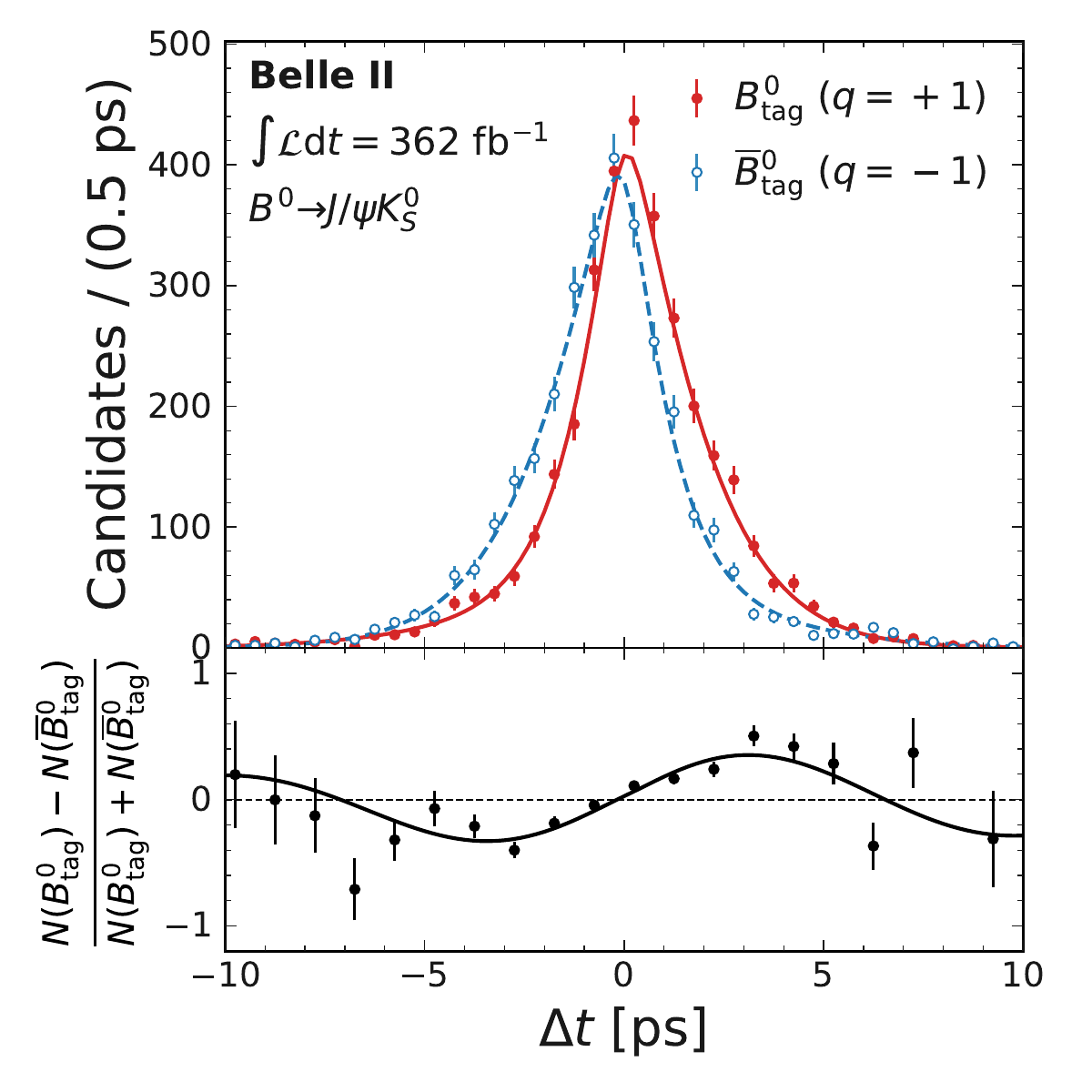}%
  \hfill%
  \includegraphics[width=\linewidth]{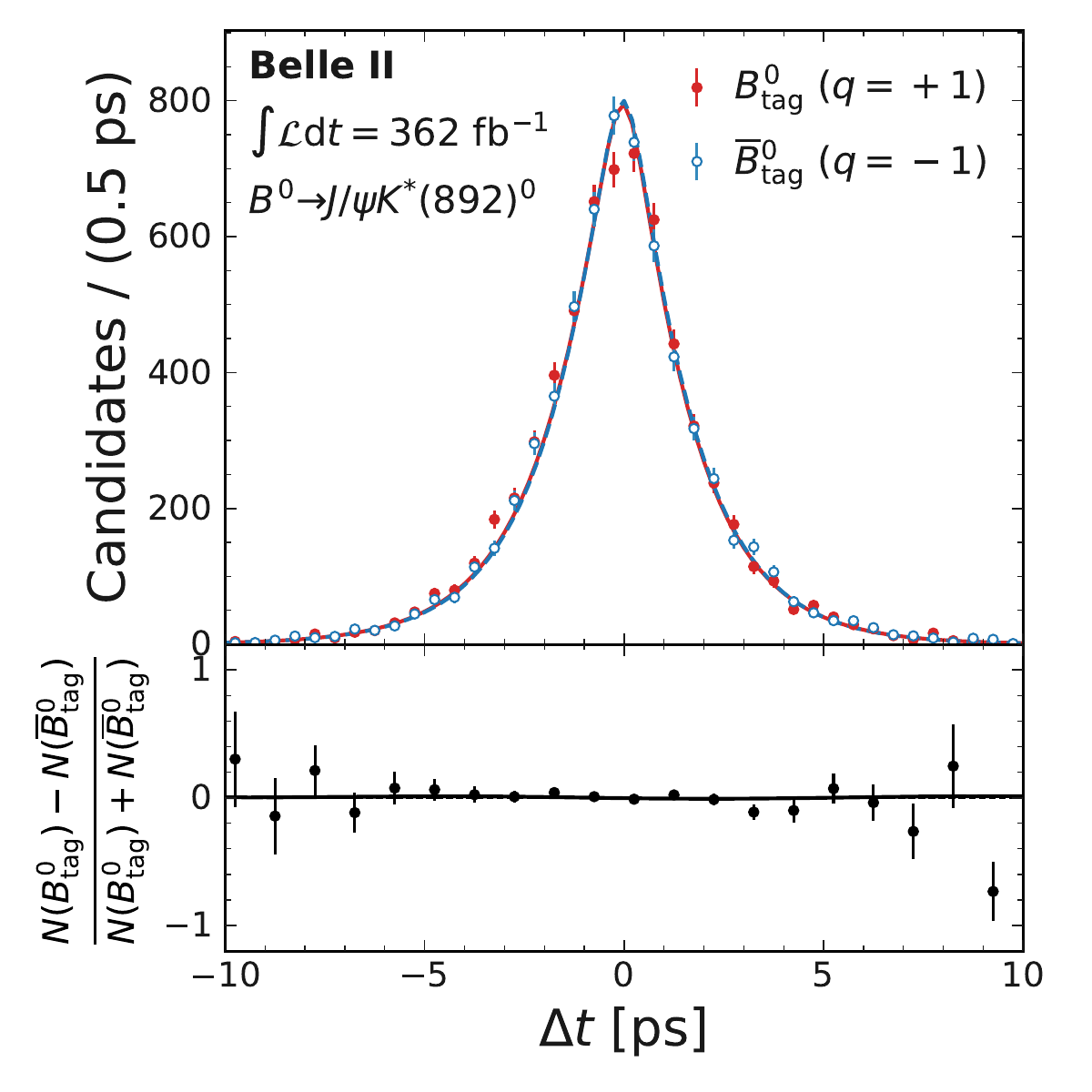}

  \caption{Background-subtracted $\Delta t$ distributions for $\PBzero\to\PJpsi\PKshortzero$~(top) and $\PBzero\to\PJpsi\PKstarzero$~(bottom) in the full $r$ range~(points) and the best-fit function~(lines) for opposite- and like-flavor \PB pairs and the corresponding asymmetries.}

  \label{fig:TDCPV_realdata}
\end{figure}

To account for resolution and bias in determining $\Delta t$, we use the resolution function of the $\PBzero\to\PDoptstarminus\Ppiplus$ decays without the outlier component, which shows no impact on the results.
The $\atag$, $\wavg$, $\Delta w$, and resolution-function parameters are fixed to the values determined from the study of $\PBzero\to\PDoptstarminus\Ppiplus$, so that the only parameters left free to vary in the $\Delta t$ fit are $S$ and $C$.
Figure~\ref{fig:TDCPV_realdata} shows the background-subtracted $\Delta t$ distributions (combining all $r$ intervals) and the result of the fits.
For $\PBzero\to\PJpsi\PKshortzero$, 
\mbox{\SCP = \SI{0.724 \pm 0.035}{}} and \mbox{\CCP = \SI{-0.035 \pm 0.026}{}}.
The statistical correlation between \SCP and \CCP is \prdcom{$-0.09$}.
For $\PBzero\to\PJpsi\PKstarzero$, 
\mbox{\SCP = \SI{-0.018 \pm 0.026}{}} and \mbox{\CCP = \SI{0.008 \pm 0.019}{}};
as expected, both are consistent with zero. The uncertainties are statistical only.

Additionally, we fit 
the $\PBzero\to\PJpsi\PKshortzero$ candidates without distinguishing
between \PBtag and \APBtag, therefore removing the ability to observe \CP violation, with $\tau$ free.
This checks for potential problems in the modeling of the resolution function, which would likely result in $\tau$ being biased from its expected value.
We measure the \PBzero lifetime to be \SI{1.514 \pm 0.022}{ps}, which agrees with the current world average~\cite{PDG}.
The uncertainty is statistical only.

Table~\ref{tab:systematics_summary} lists the statistical and systematic uncertainties on \efftag for $\PBzero\to\PDoptstarminus\Ppiplus$ and \SCP and \CCP for $\PBzero\to\PJpsi\PKshortzero$.
Statistical uncertainties are computed by bootstrapping~\cite{Efron:1979bxm}, resampling the $\PBzero\to\PDoptstarminus\Ppiplus$ and $\PBzero\to\PJpsi\PKshortzero$ data 1000 times each.
The statistical uncertainties are larger than the sum in quadrature of all the individual systematic uncertainties.

\begin{table*}[tb]
    \centering

    \caption{Systematic and statistical uncertainties on \efftag for $\PBzero\to\PDoptstarminus\Ppiplus$ and, \SCP and \CCP for $\PBzero\to\PJpsi\PKshortzero$.}

    \begin{tabular}{l*{3}{r}}

        \hline\hline

        Source
        & \efftag [\%]
        & \SCP
        & \CCP
        \\

        \hline\hline

        %%%%%
        Detector alignment
        & 0.08
        & 0.005
        & 0.003
        \\

        %%%%%
        Interaction region
        & 0.16
        & 0.002
        & 0.002
        \\

        %%%%%
        Beam energy
        & 0.03
        & $<0.001$
        & 0.001
        \\

        %%%%%
        $\Delta E$-fit background model
        & 0.11
        & 0.001
        & 0.001
        \\     

        %%%%%
        $\Delta E$-fit signal model
        & 0.08
        & 0.003
        & 0.006
        \\

        \sWeight background subtraction
        & 0.24
        & 0.001
        & 0.001
        \\

        %%%%%
        Fixed resolution-function parameters %%%
        & 0.07
        & 0.004
        & 0.004
        \\

        %%%%%
        $\tau$ and $\Delta m_{\Pdown}$
        & 0.06
        & 0.001
        & $<0.001$
        \\

        %%%%%
        $\sigma_{\Delta t}$ binning
        & 0.04
        & $<0.001$
        & $<0.001$
        \\

        %%%%%
        $\Delta t$-fit bias 
        & 0.09
        & 0.002
        & 0.005
        \\

        %%%%%
        \CP violation in \PBtag decay
        &
        & \prdcom{$<0.001$}
        & \prdcom{0.027}
        \\

        %%%%%
        $\PBzero\to\PDoptstarminus\Ppiplus$ sample size
        &
        & 0.004
        & 0.007
        \\

        \hline

        Total systematic uncertainty
        & 0.36
        & \prdcom{0.009}
        & \prdcom{0.029}
        \\

        \hline

        Statistical uncertainty
        & 0.43
        & 0.035
        & 0.026
        \\

        \hline\hline

    \end{tabular}
    \label{tab:systematics_summary}
\end{table*}

Uncertainties on the alignment of the tracking system of \belletwo detector~\cite{Bilka:2020kgr}, 
the shape and location of the IR, 
and the $\APelectron\Pelectron$ beam energy propagate to uncertainties on $\Delta t$, resulting in potential changes to \efftag, \SCP, and \CCP.
We determine \efftag, \SCP, and \CCP from simulated events reconstructed assuming four detector misalignment scenarios and take their changes, added in quadrature, as systematic uncertainties.
Both the IR and beam energy are determined from $\APelectron\Pelectron\to\APmuon\Pmuon$ events in 30-minute intervals. 
We determine \efftag, \SCP, and \CCP with the parameters of the IR and beam energy varied by their uncertainties and take the shifts as systematic uncertainties.

Uncertainties on $\Delta E$-fit component shapes propagate to uncertainties on the background-subtracted $\Delta t$ distributions, resulting in potential changes to \efftag, \SCP, and \CCP.
We fit using various models and take any resulting shifts, added in quadrature, as systematic uncertainties.
For the fit to $\PBzero\to\PDoptstarminus\Ppiplus$ data, these models are inclusion of an additional Gaussian function to model a small peaking background from $\PB\APB$ events, variation of the fixed ratio of $\PB\APB$ events to $\PBzero\to\PDoptstarminus\Ppiplus$ events by $\pm20\%$, and the freeing of the ratio of $\PBzero\to\PDoptstarminus\PKplus$ to $\PBzero\to\PDoptstarminus\Ppiplus$ events.
Variations to the background models in the fits to $\PBzero\to\PJpsi\PKshortzero$ data have negligible impact.
For the signal components, we varied fixed parameters within their uncertainties one by one.

The process of subtracting the backgrounds using \sWeight is itself a source of uncertainty.
For $\PBzero\to\PJpsi\PKshortzero$, it is accounted for in the $\Delta t$-fit bias discussed below.
We account for the uncertainty in the background subtraction in $\PBzero\to\PDoptstarminus\Ppiplus$ by determining \efftag, \SCP, and \CCP replacing the $\Delta t$ distributions with those from \SI{1}{ab^{-1}} of simulated $\PBzero\to\PDoptstarminus\Ppiplus$ data that either contain signal events or signal and background events with background subtraction using \sWeight, and take the differences as systematic uncertainties.
This is the dominant systematic uncertainty on \efftag.

Uncertainties on $\Delta t$-fit shape parameters directly propagate to changes to \efftag, \SCP, and \CCP.
We repeat the fits with fixed resolution-function parameters freed one at a time and take the resulting changes to \efftag, \SCP, and \CCP, added in quadrature, as systematic uncertainties.
We also repeat the fits with $\tau$ and $\Delta m_{\Pdown}$ varied within their known uncertainties~\cite{PDG} and take the resulting changes, added in quadrature, as systematic uncertainties.
Finally, we repeat the fits with the numbers of bins for the $\sigma_{\Delta t}$ histogrammed probability density functions varied between 200 and 1000 and take the largest changes as systematic uncertainties.

The $\Delta t$ fits have biases that we determine from fits to simulated data sets equivalent in size to the real data, 20 such sets for $\PBzero\to\PDoptstarminus\Ppiplus$ and 290 for $\PBzero\to\PJpsi\PKshortzero$.
We take the quadratic sum of the biases and their uncertainties as systematic uncertainties.

Equation~(\ref{eq:fobs}) does not account for \CP violation in \PBtag decays~\cite{Long:2003wq}. This yields a systematic uncertainty determined in Ref.~\cite{Belle:2012paq}, which is the dominant systematic uncertainty on \CCP.
\prdcom{This uncertainty can be drastically reduced by performing a combined measurement of \CCP and \SCP in \CP-odd and \CP-even decays, e.g., $\PBzero\to\PJpsi\PKshortzero$ and $\PBzero\to\PJpsi\PKlongzero$ have opposite \CP eigenvalues.}
We propagate the statistical uncertainties on \gflat's parameters and resolution-function parameters, arising from the $\PBzero\to\PDoptstarminus\Ppiplus$ sample size, to uncertainties on \SCP and \CCP by repeating the fits for each $\PBzero\to\PDoptstarminus\Ppiplus$ bootstrap sample.

%%%%%%%%%%%%%%%%%%%
\section{Summary}
\label{sec:summary}

We report on a new \PB flavor tagger, \gflat, for \belletwo that uses a graph-neural-network to account for the correlated information among the decay products of the tag-side \PB.
We calibrate it using flavor-specific hadronic \PB decays reconstructed in a \SI{362 \pm 2}{fb^{-1}} sample of \belletwo data and determine an effective tagging efficiency of
\begin{equation}
    \efftag = (37.40 \pm 0.43 \pm 0.36)\%,
\end{equation}
where the first uncertainty is statistical and the second is systematic.
For comparison, using the same data, we determine \efftag = \SI{31.68 \pm 0.45}{\percent} for the \belletwo category-based flavor tagger.\footnote{Systematic uncertainties were not explicitly computed for the category-based flavor tagger, as they are expected to be very similar to and fully correlated with those from \gflat.} 
The \gflat algorithm thus has an 18\% better effective tagging efficiency.

We demonstrate \gflat by measuring \SCP and \CCP for $\PBzero\to\PJpsi\PKshortzero$,
\begin{alignat}{5}
      \SCP &=~ && &  0.724 && ~\pm~ 0.035 && ~\pm~ \prdcom{0.009}, \\ 
      \CCP &=~ && & -0.035 && ~\pm~ 0.026 && ~\pm~ \prdcom{0.029},
\end{alignat}
with a statistical correlation between \SCP and \CCP of \prdcom{$-0.09$}.
\prdcom{This measurement supersedes our preliminary result~\cite{s2bpaper} and agrees with previous measurements~\cite{PDG, BaBar:2009byl, Belle:2012paq, LHCb:2023zcp}}.
The statistical uncertainties are 
8\% and 7\% smaller,
respectively, than they would be if measured using the category-based flavor tagger, as expected given \gflat's higher effective tagging efficiency.
From \SCP, we calculate $\phi_1$ = \SI{23.2 \pm 1.5 \pm 0.6}{\degree}.\footnote{The other solution $\pi/2-\phi_1$ is excluded from independent measurements~\cite{BaBar:2015oxm}}

\section*{Acknowledgements}
% Policy from October 20, 2022
This work, based on data collected using the Belle II detector, which was built and commissioned prior to March 2019, was supported by
%Armenia
Higher Education and Science Committee of the Republic of Armenia Grant No.~23LCG-1C011;
%Australia
Australian Research Council and Research Grants
No.~DP200101792, % Jackson
No.~DP210101900, % Urquijo
No.~DP210102831, % Sevior
No.~DE220100462, % Hsu
No.~LE210100098, % Infrastructure
and
No.~LE230100085; % Infrastructure
%Austria
Austrian Federal Ministry of Education, Science and Research,
Austrian Science Fund
No.~P~31361-N36
and
No.~J4625-N,
and
Horizon 2020 ERC Starting Grant No.~947006 ``InterLeptons'';
%Canada
Natural Sciences and Engineering Research Council of Canada, Compute Canada and CANARIE;
%China
National Key R\&D Program of China under Contract No.~2022YFA1601903,
National Natural Science Foundation of China and Research Grants
No.~11575017,
No.~11761141009,
No.~11705209,
No.~11975076,
No.~12135005,
No.~12150004,
No.~12161141008,
and
No.~12175041,
and Shandong Provincial Natural Science Foundation Project~ZR2022JQ02;
%Czech Republic
the Czech Science Foundation Grant No.~22-18469S;
%EU
European Research Council, Seventh Framework PIEF-GA-2013-622527,
Horizon 2020 ERC-Advanced Grants No.~267104 and No.~884719,
Horizon 2020 ERC-Consolidator Grant No.~819127,
Horizon 2020 Marie Sklodowska-Curie Grant Agreement No.~700525 ``NIOBE''
and
No.~101026516,
and
Horizon 2020 Marie Sklodowska-Curie RISE project JENNIFER2 Grant Agreement No.~822070 (European grants);
%France
L'Institut National de Physique Nucl\'{e}aire et de Physique des Particules (IN2P3) du CNRS
and
L'Agence Nationale de la Recherche (ANR) under grant ANR-21-CE31-0009 (France);
%Germany
BMBF, DFG, HGF, MPG, and AvH Foundation (Germany);
%India
Department of Atomic Energy under Project Identification No.~RTI 4002,
Department of Science and Technology,
and
UPES SEED funding programs
No.~UPES/R\&D-SEED-INFRA/17052023/01 and
No.~UPES/R\&D-SOE/20062022/06 (India);
%Israel
Israel Science Foundation Grant No.~2476/17,
U.S.-Israel Binational Science Foundation Grant No.~2016113, and
Israel Ministry of Science Grant No.~3-16543;
%Italy
Istituto Nazionale di Fisica Nucleare and the Research Grants BELLE2;
%Japan
Japan Society for the Promotion of Science, Grant-in-Aid for Scientific Research Grants
No.~16H03968,
No.~16H03993,
No.~16H06492,
No.~16K05323,
No.~17H01133,
No.~17H05405,
No.~18K03621,
No.~18H03710,
No.~18H05226,
No.~19H00682, % Niigata
No.~20H05850,
No.~20H05858,
No.~22H00144,
No.~22K14056,
No.~22K21347,
No.~23H05433,
No.~26220706,
and
No.~26400255,
the National Institute of Informatics, and Science Information NETwork 5 (SINET5), 
and
the Ministry of Education, Culture, Sports, Science, and Technology (MEXT) of Japan;  
%Korea
National Research Foundation (NRF) of Korea Grants
No.~2016R1\-D1A1B\-02012900,
No.~2018R1\-A2B\-3003643,
No.~2018R1\-A6A1A\-06024970,
No.~2019R1\-I1A3A\-01058933,
No.~2021R1\-A6A1A\-03043957,
No.~2021R1\-F1A\-1060423,
No.~2021R1\-F1A\-1064008,
No.~2022R1\-A2C\-1003993,
and
No.~RS-2022-00197659,
Radiation Science Research Institute,
Foreign Large-Size Research Facility Application Supporting project,
the Global Science Experimental Data Hub Center of the Korea Institute of Science and Technology Information
and
KREONET/GLORIAD;
%Malaysia
Universiti Malaya RU grant, Akademi Sains Malaysia, and Ministry of Education Malaysia;
%Mexico
% CINVESTAV-IPN, UNAM, UAS, BUAP and CONACYT are funded under
Frontiers of Science Program Contracts
No.~FOINS-296,
No.~CB-221329,
No.~CB-236394,
No.~CB-254409,
and
No.~CB-180023, and SEP-CINVESTAV Research Grant No.~237 (Mexico);
%Poland
the Polish Ministry of Science and Higher Education and the National Science Center;
%Russia
the Ministry of Science and Higher Education of the Russian Federation
and
the HSE University Basic Research Program, Moscow;
%Saudi Arabia
University of Tabuk Research Grants
No.~S-0256-1438 and No.~S-0280-1439 (Saudi Arabia);
%Slovenia
Slovenian Research Agency and Research Grants
No.~J1-9124
and
No.~P1-0135;
%Spain
Agencia Estatal de Investigacion, Spain
Grant No.~RYC2020-029875-I
and
Generalitat Valenciana, Spain
Grant No.~CIDEGENT/2018/020;
%Taiwan
National Science and Technology Council,
and
Ministry of Education (Taiwan);
%Thailand
Thailand Center of Excellence in Physics;
%Turkey
TUBITAK ULAKBIM (Turkey);
%Ukraine
National Research Foundation of Ukraine, Project No.~2020.02/0257,
and
Ministry of Education and Science of Ukraine;
%USA
the U.S. National Science Foundation and Research Grants
No.~PHY-1913789 % Indiana CEEM
and
No.~PHY-2111604, % Luther
and the U.S. Department of Energy and Research Awards
No.~DE-AC06-76RLO1830, % PNNL
No.~DE-SC0007983, % Wayne State
No.~DE-SC0009824, % Florida
No.~DE-SC0009973, % VPI
No.~DE-SC0010007, % Duke
No.~DE-SC0010073, % South Carolina
No.~DE-SC0010118, % Carnegie Mellon
No.~DE-SC0010504, % Hawaii
No.~DE-SC0011784, % Cincinnati
No.~DE-SC0012704, % BNL
No.~DE-SC0019230, % Duke
No.~DE-SC0021274, % Mississippi
No.~DE-SC0021616, % Mississippi
No.~DE-SC0022350, % Louisville
No.~DE-SC0023470; % South Alabama
%last group
and
%Vietnam
the Vietnam Academy of Science and Technology (VAST) under Grants
No.~NVCC.05.12/22-23
and
No.~DL0000.02/24-25.

% Policy from October 20, 2022
These acknowledgements are not to be interpreted as an endorsement of any statement made
by any of our institutes, funding agencies, governments, or their representatives.

We thank the SuperKEKB team for delivering high-luminosity collisions;
the KEK cryogenics group for the efficient operation of the detector solenoid magnet;
the KEK computer group and the NII for on-site computing support and SINET6 network support;
and the raw-data centers at BNL, DESY, GridKa, IN2P3, INFN, and the University of Victoria for off-site computing support.

\appendix

\section{\gflat parameters}
\label{sec:app_params}

\begin{figure*}[htbp]
    \centering
    \includegraphics[width=0.8\textwidth]{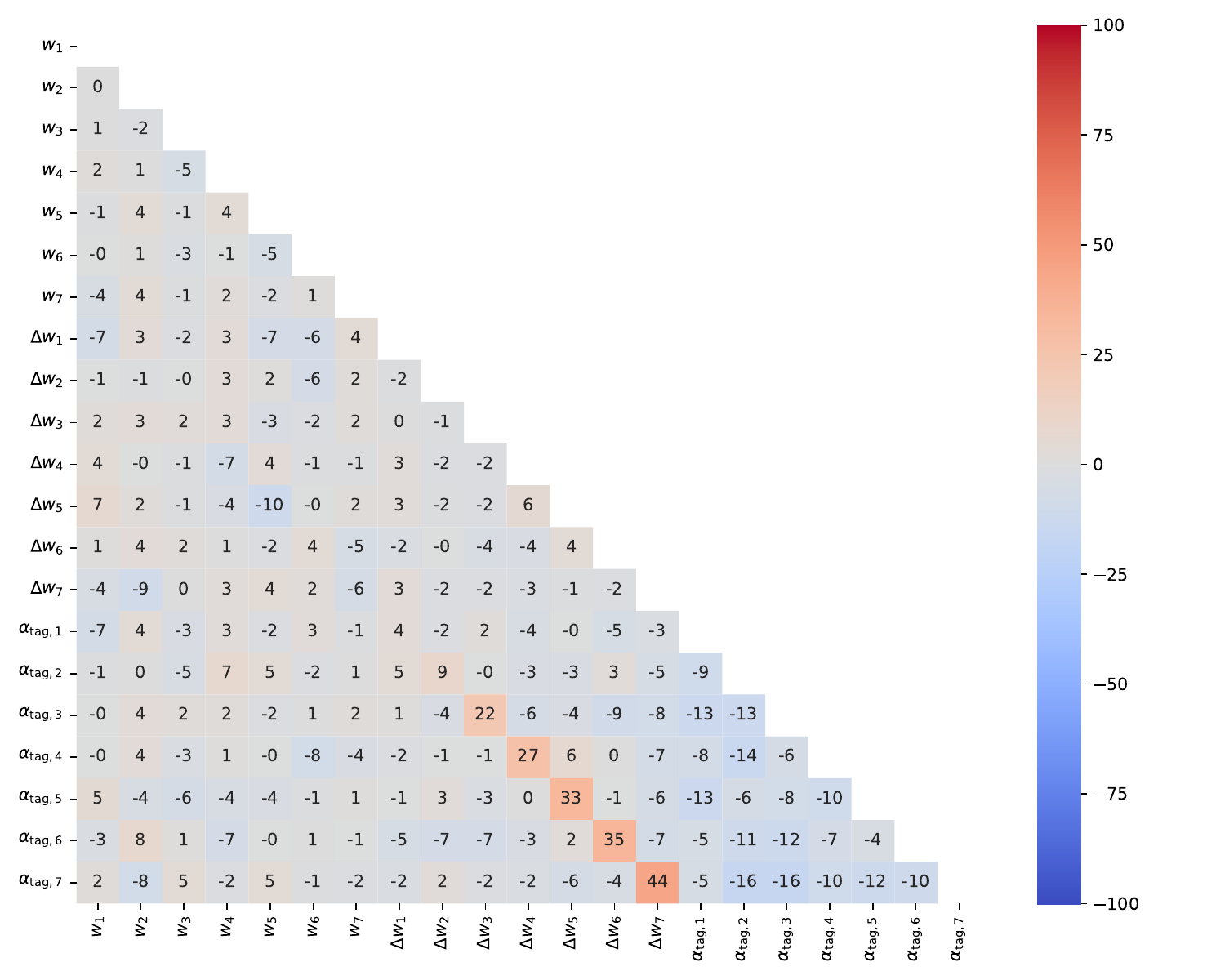}
    \caption{Correlation coefficients, in $10^{-2}$, between the \gflat parameters.
    Subscripts indicate $r$ bins.}
    \label{fig:FTreso_params_corrs_reduced}
\end{figure*}

Table~\ref{tab:data_fitted_parameters} lists \atag, \wavg, and $\Delta w$ for each $r$ bin, measured from events with $\PBzero\to\PDoptstarminus\Ppiplus$. 
The sources of systematic uncertainty are the same as listed in Table~\ref{tab:systematics_summary} for \efftag.
Figure~\ref{fig:FTreso_params_corrs_reduced} shows the statistical correlation coefficients  between the parameters that are used as inputs to estimate systematic uncertainties for \SCP and \CCP.

\begin{table*}[htbp]
    \centering

    \caption{\gflat parameters in each $r$ bin.}

    \label{tab:data_fitted_parameters}

    \begin{tabular}{l*{3}{r}}

        \hline
        \hline
    
        $r$ bin
        & \atag [\%]
        & \wavg [\%]
        & $\Delta w$ [\%]
        \\

        \hline
        
        \numrange{0.0}{0.1}%1
        & $-1.72 \pm 1.47 \pm 1.32$
        & $48.29 \pm 0.78 \pm 0.75$
        & $0.78 \pm 1.16 \pm 0.71$
        \\
        
        \numrange{0.1}{0.25}%2
        & $-0.94 \pm 1.36 \pm 1.45$
        & $42.07 \pm 0.72 \pm 0.32$
        & $-1.41 \pm 1.06 \pm 0.92$
        \\
        
        \numrange{0.25}{0.45}%3
        & $-0.28 \pm 1.28 \pm 1.46$
        & $34.63 \pm 0.61 \pm 0.61$
        & $-0.04 \pm 0.97 \pm 1.28$
        \\
        
        \numrange{0.45}{0.6}%4
        & $3.21 \pm 1.44 \pm 1.50$
        & $24.17 \pm 0.68 \pm 0.36$
        & $1.64 \pm 1.13 \pm 0.52$
        \\
        
        \numrange{0.6}{0.725}%5
        & $1.17 \pm 1.58 \pm 1.47$
        & $16.98 \pm 0.68 \pm 0.92$
        & $1.36 \pm 1.15 \pm 0.72$
        \\
        
        \numrange{0.725}{0.875}%6
        & $-1.13 \pm 1.30 \pm 1.55$
        & $11.50 \pm 0.53 \pm 0.39$
        & $-0.26 \pm 0.92 \pm 0.71$
        \\
        
        \numrange{0.875}{1.0}%7
        & $-0.18 \pm 0.91 \pm 1.30$
        & $2.62 \pm 0.27 \pm 0.14$
        & $0.75 \pm 0.53 \pm 0.60$
        \\

        \hline
        \hline

    \end{tabular}
\end{table*}

\newpage
\bibliography{./references}

\end{document}